\documentclass[12pt,a4paper]{article}
\usepackage{setspace}
\usepackage{a4wide}
\usepackage{amssymb}
\usepackage{amsmath}
\usepackage{graphicx}
\usepackage{cancel}
\usepackage{array}
\usepackage{epsfig} 
\usepackage{hyperref}
\newcolumntype{x}[1]{%
>{\centering\hspace{0pt}}m{#1}}%
\usepackage{ulem}
\usepackage{tikz}
\usetikzlibrary{decorations}
\usetikzlibrary{decorations.pathmorphing}
\usetikzlibrary{decorations.markings}
\usetikzlibrary{patterns}
\usepackage{slashed}
\newcolumntype{C}{>{\centering\arraybackslash}m{4cm}}
\newcommand{\beq}{\begin{equation}}
\newcommand{\eeq}{\end{equation}}
\newcommand{\beqa}{\begin{eqnarray}}
\newcommand{\eeqa}{\end{eqnarray}}
\newcommand{\nn}{\nonumber}
\newcommand{\Tr}{\textrm{tr}}
\newcommand{\zb}{\bar{z}}
\newcommand{\psib}{\bar{\psi}}
\newcommand{\Lag}{\mathcal{L}}
\newcommand{\pa}{\partial}
\newcommand{\D}{D}

\newcommand{\N}{N}
\newcommand{\e}{k}
\newcommand{\p}{p}
\newcommand{\Op}{O}
\newcommand{\ka}{\kappa}
\newcommand{\amp}{0.5mm}
\newcommand{\wid}{0.5pt}
\newcommand{\ft}[2]{{\textstyle\frac{#1}{#2}}}
\begin{document}
%\usetikzlibrary{decorations}
\thispagestyle{empty}
\onehalfspace

\setcounter{footnote}{0}

\begin{center}
{\Large{\bf On the integrability of two-dimensional models\\
\vspace{2mm}
 with $U(1)\times SU(\N)$ symmetry}}
\vspace{15mm}

{\sc Benjamin Basso$^a$, Adam Rej$^{b,c}$} \\[5mm]

{\it $^a$ Princeton Center for Theoretical Science, Jadwin Hall, Princeton University, Princeton, NJ 08544, USA}\\[5mm]

{\it $^b$  School of Natural Sciences, Institute for Advanced Study, Princeton, NJ 08540, USA}\\[2mm]

{\it $^c$  Marie Curie Fellow}\\[5mm]

\texttt{\big\{bbasso $\arrowvert$ arej\big\} @ $\left\{\frac{\texttt{\normalsize princeton.edu}}{\texttt{\normalsize ias.edu}} \right\}$}
\\[15mm]

\textbf{Abstract}\\[2mm]
\end{center}

\noindent{In this paper we study the integrability of a family of models with $U(1)\times SU(\N)$ symmetry. They admit fermionic and bosonic formulations related through bosonization and subsequent T-duality. The fermionic theory is just the $CP^{\N-1}$ sigma model coupled to a self-interacting massless fermion, while the bosonic one defines a one-parameter deformation of the $O(2\N)$ sigma model. For $\N=2$ the latter model is equivalent to the integrable deformation of the $O(4)$ sigma model discovered by Wiegmann. At higher values of $\N$ we find that integrability is more sporadic and requires a fine-tuning of the parameters of the theory. A special case of our study is the $\N=4$ model, which was found to describe the $AdS_4\times CP^3$ string theory in the Alday-Maldacena decoupling limit. In this case we propose a set of asymptotic Bethe ansatz equations for the energy spectrum.}

\newpage
\tableofcontents
\newpage

\section{Introduction}

Two-dimensional theories are physically interesting. They are easier to work with than the four-dimensional ones yet they exhibit phenomenons like confinement, charge screening, dynamical transmutation, etc. A prominent example is the so-called $CP^{\N-1}$ sigma model~\cite{Eichenherr:1978qa, D'Adda:1978uc, Witten:1978bc} that mimics and simplifies many of the expected features of four-dimensional confining gauge theories. Microscopically the model is defined by the Lagrangian
\beq \label{CPn}
\Lag = \ka(\pa_{\mu}-iA_\mu)\bar{z} (\pa^{\mu}+iA^\mu) z \,,
\eeq
where $z = (z_1, \ldots , z_{\N})$ is an $SU(\N)$ multiplet of complex bosons subject to the constraint $\bar{z}z =1$ and $A_\mu$ is an abelian gauge field. It has $U(1)$ gauge symmetry $A_{\mu} \rightarrow A_{\mu} - \pa_\mu \omega, z\rightarrow e^{i\omega}z$, but no kinetic term for the gauge field. The latter may be integrated out and replaced by a local interaction $A_\mu = i \bar{z}\pa_\mu z$. The model~(\ref{CPn}) has a single dimensionless coupling $\sim 1/\kappa$, it is known to be renormalizable~\cite{Valent:1984rj, Hikami:1979ih} and is asympotically free. To the leading order $\kappa(\mu) \sim \beta_0 \log{(\mu/\Lambda)}$ with $\Lambda$ being the dynamical scale and $\beta_0 = \N/2\pi$. Similarly to the $O(2\N)$ model \cite{Polyakov:1975rr, Polyakov:1987ez}, the model \eqref{CPn} is solvable at large $\N$, see \cite{D'Adda:1978uc, Witten:1978bc}, and the fundamental excitations acquire a mass $\sim \Lambda$. What makes it different, however, is that the fundamental excitations are confined by the long-range Coulomb interaction induced non-perturbatively by the gauge field. The spectrum has a gap and is populated by ``mesons'' falling into representations of $SU(\N)/\mathbb{Z}_{\N}$, with $\mathbb{Z}_{\N}$ the center of $SU(\N)$. 

Much less is known about the $CP^{\N-1}$ model beyond the large $\N$ limit. Classically the model has infinitely many conserved currents and is completely integrable \cite{Eichenherr:1978qa}. Unfortunately, these conservation laws are spoiled by anomalies at the quantum level~\cite{Abdalla:1980jt} and the integrability is not preserved by the quantization. A noticeable exception occurs for $\N=2$ when the theory is equivalent to the $O(3)$ sigma model~\cite{D'Adda:1978uc, Witten:1978bc} and hence integrable~\cite{Pohlmeyer:1975nb, Zamolodchikov:1977nu, Polyakov:1977vm, Zamolodchikov:1978xm}. A more promising class of models from the perspective of integrability is obtained by minimally coupling a massless Dirac fermion to the gauge theory  \eqref{CPn}. The corresponding Lagrangian is
\beq \label{fermionicmodels}
\Lag = \ka(\pa_{\mu}-iA_\mu)\bar{z} (\pa^{\mu}+iA^\mu) z + i\psib\gamma^{\mu}(\pa_{\mu}-i\e A_\mu)\psi -\frac{\lambda}{2} \left(\psib \gamma_\mu \psi \right)^2\,,
\eeq 
where the fermion has charge $\e$ and a self-interaction controlled by the Thirring coupling $\lambda$.  As opposed to \eqref{CPn} the theory \eqref{fermionicmodels} has an additional $U(1)$ symmetry associated to the conservation of the  number of fermions. At the classical level it also has an axial $U(1)$ symmetry coming from chiral rotations of the fermion. This symmetry becomes anomalous at the quantum level and is broken down to a discrete $\mathbb{Z}_{2\e}$ subgroup, which in turn is spontaneously broken down to $\mathbb{Z}_2$~\cite{Witten:1978bc}.

From the viewpoint of perturbation theory both models \eqref{CPn} and \eqref{fermionicmodels} look similar. For instance, to leading order at weak coupling the running of the coupling $1/\kappa$ is oblivious to the presence of the fermion and the theory \eqref{fermionicmodels} is equipped with the dynamical scale $\Lambda$ of theory \eqref{CPn}. The most striking difference between the models \eqref{CPn} and \eqref{fermionicmodels} is non-perturbative. As evidenced by the large $\N$ analysis~\cite{Witten:1978bc, D'Adda:1978kp} the $U(1)$ gauge symmetry is spontaneously broken in \eqref{fermionicmodels}. The phenomenon is similar to the one observed in the Schwinger model; the massless fermion is eaten up by the gauge field which in turn acquires a mass of order $\sim \e\Lambda/\sqrt{N (1 + \lambda/\pi)}$ for large $\N$ and becomes dynamical. This leads to the screening of the Coulomb interaction at long distance and the fundamental excitations, henceforth dubbed spinons, are liberated. A natural question in these new circumstances is whether the system can now become integrable.

In this paper we will give a positive answer to this question. We will argue that for any value of $\N$ and the fermion charge $\e$ there exists one value of the Thirring coupling, or more accurately one renormalization group (RG) trajectory, for which the system~(\ref{fermionicmodels}) is integrable and described by the minimal reflectionless $U(\N)$ S-matrix. The latter S-matrix corresponds to the class II solution in the classification of $U(\N)$ invariant factorized S-matrix established by Berg et al.~\cite{Berg:1977dp}.

We should stress at this point that our discussion is not unrelated to earlier considerations. It has been known for a long time that the integrability of the quantum $CP^{\N-1}$ model can be restored by adding  massless Dirac fermions \cite{Koberle:1982ju, Koberle:1987wc}. Such a model was proposed by K\"oberle and Kurak \cite{Koberle:1987wc} who introduced $\N$ Dirac fermions with charge $\e=1$. The coupling to the bosonic degrees of freedom was chosen to be minimal and the Thirring coupling was set to zero. To verify their proposal the authors of \cite{Koberle:1987wc} constructed the S-matrix to leading order at large $\N$ and observed it to agree with the large $\N$ expansion of the minimal reflectionless S-matrix of \cite{Berg:1977dp}. Further support for the integrability of this model was given in~\cite{Abdalla:1981yt, Abdalla:1985nm}.

The model which we study in this article is not essentially different from the one considered by K\"oberle and Kurak in~\cite{Koberle:1987wc}. It has however couple of advantages over the latter. First of all, theory \eqref{fermionicmodels} is minimal because it contains only \textit{one} massless Dirac fermion. It is known that only one fermion is effectively active even if multiple massless fermions are minimally coupled to a $U(1)$ gauge field in two dimensions \cite{Callan:1977qb}. This was also observed by the authors of \cite{Koberle:1982ju, Koberle:1987wc}. From this point of view the model \eqref{fermionicmodels} appears as the most generic effective theory. Furthermore, the theory \eqref{fermionicmodels} is renormalizable, at one-loop at least, as opposed to models without Thirring coupling. We shall prove that the Thirring coupling $\lambda$ in \eqref{fermionicmodels} is not exactly marginal and starts running at the one-loop level
\beq\label{ThIntro}
\lambda = \lambda_\infty - {\e^2\over 2\N\kappa} + O(1/\kappa^2)\,.
\eeq
The parameter $\lambda_{\infty}$ denotes the UV value of the Thirring coupling $\lambda$. It is physical and may be chosen arbitrarily. Consistency with the factorized S-matrix requires the Thirring coupling to run and the parameter $\lambda_{\infty}$ to be fine-tuned. We shall show this by studying the free energy density of a gas of spinons at large chemical potential. The computation may be done either by using the conjectured exact S-matrix or with help of the standard perturbation theory. Matching both calculations is very sensitive to the renormalization properties of the Thirring coupling and confirms~(\ref{ThIntro}). The theory \eqref{fermionicmodels} may thus be seen to provide a refined version of the model introduced by K\"oberle and Kurak such that it is consistent with the usual requirements of a perturbative QFT.
 
Another motivation for studying the theory~(\ref{fermionicmodels}) is tied to the AdS/CFT correspondence. A model of this class has been recently found by Bykov~\cite{Bykov:2010tv} to describe the truncation of the $AdS_4 \times CP^3$ type IIA super-string sigma in the Alday-Maldacena decoupling limit~\cite{Alday:2007mf}. The latter is understood as the low-energy limit of the string sigma model in the background of a long string rotating in $AdS_3 \subset AdS_4$ \cite{Gubser:2002tv}. The effective  Lagrangian reads explicitly~\cite{Bykov:2010tv}
\beq \label{Bykovsmodel}
\Lag =\ka(\pa_{\mu}-iA_\mu)\bar{z} (\pa^{\mu}+iA^\mu) z + i\psib\gamma^{\mu}(\pa_{\mu}-2iA_\mu)\psi + \frac{1}{4\ka} \left(\psib \gamma_\mu \psi \right)^2\,,
\eeq
with the bosonic variables describing $CP^3$ and the massless Dirac fermion being the only extra remnant of the superstring coordinates in the massless limit~\cite{Alday:2008ut}. The effective model \eqref{Bykovsmodel} belongs to the class \eqref{fermionicmodels} with $\N=4$, $\e=2$, and $\lambda_{\infty}=0$. 

It is of particular interest to study the integrability of this effective model. The sigma model on $AdS_4 \times CP^3$, analogously to the $AdS_5 \times S^5$ case, is believed to be quantum integrable. We refer the reader to \cite{Klose:2010ki} for a recent review. The amount of evidence supporting this supposition is significantly smaller than for the $AdS_5 \times S^5$ theory. In this paper we will give a non-perturbative argument in favor of the integrability of the \textit{full} string sigma model by showing that the model \eqref{Bykovsmodel} belongs to the integrable subclass of~\eqref{fermionicmodels}. We shall further propose a set of asymptotic Bethe ansatz (ABA) equations for the energy levels of the theory~\eqref{Bykovsmodel} in finite volume, i.e., defined on a cylinder of length $L$, and discuss subtle effects related to the restoration of the $\mathbb{Z}_{\e} \simeq \mathbb{Z}_{2\e}/\mathbb{Z}_{2}$ chiral symmetry. We will see that these finite-volume effects require a twisting of the ABA equations. Eventually, it should be possible to compare our results with the predictions coming from the conjectured all-loop ABA equations \cite{Gromov:2008qe, Minahan:2008hf} for the Aharony-Bergman-Jafferis-Maldacena (ABJM) theory \cite{Aharony:2008ug}. This comparison is beyond the scope of our analysis but would provide a stringent all-order test of this conjecture. 

Last but not least, we shall show that the fermionic model \eqref{fermionicmodels} is \textit{equivalent} to
\beq \label{Lagrangian}
\Lag={R^2 \over 4\pi}(\pa_\mu -iA_{\mu}) \zb (\pa^{\mu}+iA^{\mu})z + {r^2 \over 4\pi(1 -r^2/R^2)} A_{\mu}A^{\mu}\, ,
\eeq
which defines a one-parameter family of marginally relevant deformations of the $O(2\N)$ sigma model preserving the $U(1)\times SU(\N)$ symmetry. First indications of this duality were given in \cite{Campostrini:1993fr} where the fermionic model considered by K\"oberle and Kurak was related to the bosonic model~(\ref{Lagrangian}) at the level of their large $\N$ effective actions. In this article we shall establish the equivalence between \eqref{fermionicmodels} and \eqref{Lagrangian} by means of bosonization and subsequent T-dualization. This will lead to the following identification of the parameters
\beq\label{Intro:coup}
R^2 = 4\pi \kappa\, , \qquad r^{2} = {2 \e^2 \over 1+\lambda'/\pi}\, , \qquad \lambda' = \lambda + {\e^2\over 2\kappa}\, ,
\eeq
while the charge $\e$ will be found to parametrize a $\mathbb{Z}_\e$ quotient of the model \eqref{Lagrangian}. Geometrically the deformation of the sphere $S^{2\N-1}$ described by the sigma model \eqref{Lagrangian} is also known as a squashing, with the round sphere being recovered for  $r^2=R^2$. Notice also that in the bosonic formulation \eqref{Lagrangian} there is no gauge symmetry, due to the ``mass term'' $\sim A_\mu A^{\mu}$. For this reason the model is sometimes termed the ``massive'' $CP^{\N-1}$ model \cite{Azaria:1995wg}. It is nevertheless more suitable to view this model as a deformation of the $O(2\N)$ sigma model since it entails $2\N-1$ massless fields in the UV, that is one more than there are available in the case of the $CP^{\N-1}$ model. The latter is recovered in the limit $r^2 \sim 0$ when one degree of freedom decouples.

In the special case of $\N=2$, the sigma model \eqref{Lagrangian} is equivalent to the one parameter deformation of the $O(4)$ sigma model introduced by Wiegmann and Polyakov~\cite{Wiegmann:1985jt, Polyakov:1983tt}. This model is well-studied and was found to be integrable both at the classical and quantum level~\cite{Wiegmann:1985jt, Cherednik, Balog:1999ik, Balog:2000wk, Forgacs:2000eu, Kawaguchi:2010jg, Kawaguchi:2011ub, Kawaguchi:2011pf, Kawaguchi:2012ve}. Since it belongs to the family of models considered here, it offers valuable insights into the non-perturbative physics at smallest non-trivial value of $\N$. This theory remains nonetheless exceptional from the view point of integrability. Most of the arguments regarding the generic $\N$ case put forward in this paper hold for $\N > 2$ and for each of them we will point out what makes the case $\N=2$ so special. 

This paper is structured as follows. In Section~\ref{TM} we present the model in its two formulations and relate them via bosonization and T-duality. We also discuss their renormalizability and infrared physics. The classical and quantum integrability of the model are subject of Section~\ref{Int}. In Section~\ref{MBA} we perform a direct test of the proposed S-matrix at finite values of $\N$ by computing the free energy density of the model at finite chemical potential. In Section~\ref{BSM} we apply these results to the particular case relevant to string theory and construct the associated ABA equations. Section~\ref{Concl} contains concluding remarks. Technical details of our analysis are deferred to the various appendices.

\section{The models}\label{TM}

In this section we introduce the fermionic and bosonic formulations, discuss their renormalizability and establish their equivalence. We start with the bosonic model whose physics may be more easily exposed.

\subsection{The bosonic model}

The bosonic model is defined by the Lagrangian~\eqref{Lagrangian}, with the $\N$ complex scalar fields $z = (z_1, \ldots, z_\N)$ spanning the unit sphere $S^{2\N-1}$, i.e., satisfying $\zb z =1$. An equivalent formulation is obtained if one decides to integrate out the dummy field $A_{\mu}$. One finds
\beq\label{BM}
\Lag = {R^2\over 4\pi} (\partial_{\mu}-iB_{\mu})\bar{z} (\partial^{\mu}+iB^{\mu})z + {r^2\over 4\pi}B_{\mu}B^{\mu}\, ,
\eeq
where $B_\mu = i\bar{z}\partial_{\mu}z$. The first term in~(\ref{BM}) has a local $U(1)$ symmetry and coincides with the Lagrangian of the $CP^{\N-1}$ model. The last term breaks this gauge symmetry down to the global $U(1)$ subgroup and geometrically describes a circle of radius $r$ fibered over $CP^{\N-1}$. For generic values of $r$ the model~(\ref{BM}) has thus $U(1)\times SU(\N)$ symmetry. The symmetry is enhanced if the two radii are equal $r=R$ when the model becomes equivalent to the $O(2\N)$ sigma model. In the limit $r\rightarrow 0$ the small circle shrinks to a point and the theory reduces to the $CP^{\N-1}$ sigma model. The regime considered in this paper is bounded by these two special values $0 \leqslant r \leqslant R$.

The Lagrangian~(\ref{BM}) makes the geometry of the problem manifest and will be the starting point for performing the T-duality in the next subsection. It is nonetheless entirely equivalent to
\beq \label{BMdef}
\Lag = \kappa\partial_{\mu}\bar{z} \partial^{\mu}z + \kappa \eta (\zb\partial_{\mu} z)^2\,,
\eeq
where $\eta$ and $\kappa$ are dimensionless parameters, related to the radii by $R^2 = 4\pi \kappa\, , r^2 = 4\pi \kappa(1-\eta)$. This form makes the relation to the $O(2\N)$ model evident and is sometimes more convenient for semiclassical computation. Notice that for the afore-stated domain of $R$ the parameter $\eta$ takes values in the interval $0 \leqslant \eta \leqslant 1$. The boundary values $\eta = 0, 1$ correspond to the $O(2\N)$ and the $CP^{\N-1}$ model, respectively. 

\begin{figure}
\begin{center}
\includegraphics[scale=1.4]{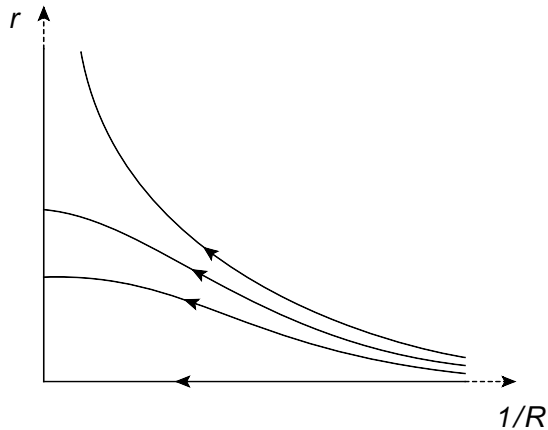}
\end{center}
\caption{A specimen of the solutions to the RG equations in the asymptotically free domain $0 \leqslant r \leqslant R$. The upper line correponds to the $O(2\N)$ trajectory $r=R$ and the lower one to $CP^{\N-1}$ with $r=0$. The arrows point toward increasing values of the renormalization scale $\mu$. All trajectories in the band $0\leqslant r < R$ end on the line $1/R=0$ at $\mu = \infty$.} \label{fig:specimen}
\end{figure}

As in the $O(2\N)$ sigma model, the radii start running at the quantum level. To reveal this effect one should proceed with the renormalization of the model~(\ref{BM}). This analysis can be done perturbatively as an expansion at large $R$ and was performed in detail in \cite{Azaria:1995wg,Balog:1999ik}. The theory was found to be asymptotically free in the domain $0 \leqslant r \leqslant R$ and the RG equations were constructed up to two loops. They take the following form
\beq\label{RGEBM}
\mu{\partial R^2 \over \partial \mu} =  2\N + {2(4\N-r^2) \over R^2} + \Op(1/R^4)\, , \qquad \mu{\partial r^2 \over \partial \mu} =  {2(\N-1) r^4\over R^4} + \Op(1/R^8)\, ,
\eeq
for $R \gg 1$ and fixed $r$. Both radii are monotonically increasing functions of the renormalization scale $\mu$, but while $R$ grows without any bound, the radius $r$ reaches a finite value in the UV. Put differently, in the $(r, R)$ plane these equations describe a flow of trajectories ending on a line of fixed points at $R=\infty$, see Figure \ref{fig:specimen}. This line is parameterized by the UV value $r_\infty = r(\infty)$ of the radius $r$, which by definition is RG-invariant. The second RG-invariant is the dynamical scale $\Lambda$ which controls the running of the radii. Explicitly, the solution to \eqref{RGEBM} is
\beq\label{R2}
%\begin{aligned}
R^2 = 2\N\log{(\mu/\Lambda)} +2(2-p)\log{\log{(\mu/\Lambda)}}+ o(1)\, ,
\eeq
and
\beq\label{r2}
\frac{1}{r^2} = \frac{1}{2Np} + {\N-1 \over \N R^2} -{(\N-1)(2-p) \over \N R^4} + \Op(1/R^6)\, .
%\end{aligned}
\eeq
Here we have introduced the parameter 
\beq \label{bosonicp}
\p = \frac{r^2_{\infty}}{2\N}\,,
\eeq
which will turn out to be convenient when discussing the physical properties of the model. Please note that the point $\p=0$ corresponds to the $CP^{\N-1}$ sigma model. The $O(2N)$ sigma model, on the other hand, is formally recovered when $\p \to \infty$.

What can one infer about the non-perturbative infrared features of the model? A standard method to derive hints about the physics beyond the perturbation theory is the large $\N$ expansion. This study has been conducted for the model at hand in~\cite{Campostrini:1993fr, Azaria:1995wg} and closely parallels the one for the dual fermionic model~\cite{Witten:1978bc, D'Adda:1978kp}. For the sake of the large $\N$ analysis, it is convenient to introduce two auxiliary fields: the gauge field $A_\mu$ as in \eqref{Lagrangian} and a scalar field $\alpha$ which implements the constraint $\zb z =1$. As in the case of the $O(2\N)$ sigma model, it is found that $\alpha$ acquires a vacuum expectation value giving a mass $m= \Lambda$ to the $2\N$ particles. These are our spinons and antispinons transforming in the fundamental and antifundamental representation of $SU(N)$, respectively. They have opposite charge with respect to the $U(1)$ group. The interactions, which are weak at large $\N$, are mediated by the gauge and the scalar field. The vector interactions play a prominent role in the physics of the model and distinguish this theory from the $O(2\N)$ sigma model. They are controlled by the gauge field propagator
\beq \label{gpropagator}
\left<A^\mu A^\nu\right> (k^2) = {-i\pi \over \N \left(A(k^2)-\p \right)}\left(\eta^{\mu \nu}-{k^{\mu}k^{\nu} \over k^2}\right) + {i\pi k^{\mu}k^{\nu} \over \N \p k^2}\, ,
\eeq
with the function $A(k^2)$ defined for $k^2 < 0$ by
\beq
A(k^2) = 1 - {1\over 2}\sqrt{1-{4m^2 \over k^2}}\textrm{arccosh}{\left(1-{k^2 \over 2m^2}\right)} = {k^2 \over 12m^2} + \Op(k^4/m^4)\,,
\eeq
and otherwise by analytic continuation. Its structure is reminiscent of the Proca propagator for a massive gauge field with mass term controlled by the parameter $\p$. For $\p=0$, when the model is equivalent to $CP^{\N-1}$, it has a pole at $k^2=0$ leading to a linear confining potential. The spectrum consists of mesons, which are spinon-antispinon bound states~\cite{D'Adda:1978uc, Witten:1978bc}. For $0<\p<1$ there is a pole in the propagator~(\ref{gpropagator}) at finite value of $k^2$. The force is short-ranged and the spinons are liberated. As long as $\p$ is very small in value, the mass of these particles is still much larger than that of the gauge field, which can be viewed as a singlet bound state. A transition occurs at $\p=1$ when the lightest bound states, including the gauge field, hit the two-particle threshold. This point is special from the integrability perspective, as already pointed out in~\cite{Campostrini:1993fr} building on~\cite{Koberle:1987wc}, and we shall come back to it later on. Finally, the range $\p\geqslant 1$ corresponds to repulsive interactions and no bound states. We refer the reader to~\cite{Campostrini:1993fr} for a more detailed discussion of the large $\N$ theory.

\subsection{T-duality}

To unveil the relation between the bosonic and the fermionic theory we need to T-dualize the former along its fibered circle.%
\footnote{We are most grateful to D.~Andriot for helpul discussions on this topic.} To perform this transformation it is easier to work with the Lagrangian~(\ref{BM}) and use local coordinates which trivialize the $U(1)$ bundle. One can write for instance $z_i=e^{i\vartheta}z'_i$ where $\vartheta$ is a compact boson with period $2\pi$, $\vartheta \sim \vartheta + 2\pi$, and $z'_i$ is a set of local coordinates on $CP^{\N-1}$. In these coordinates the one-form $B_\mu$ decomposes into vertical and horizontal parts,
\beq\label{Btb}
B_{\mu} = -\partial_{\mu}\vartheta + b_{\mu}\, ,
\eeq
where $b_\mu$ is a local one-form on $CP^{\N-1}$. In this formulation the  $U(1)$ symmetry is implemented as translations of $\vartheta$ and this isometry allows us to T-dualize along this direction. Notice in this regard that all the dependence on $\vartheta$ resides in the second term in~(\ref{BM}) and is manifest in~(\ref{Btb}). To perform the T-duality we will therefore focus only on this part of the Lagrangian, which we denote as $\Lag_{fiber}$.

To accommodate for the fermion charge $\e$, one actually needs to consider a $\mathbb{Z}_\e$ quotient of the bosonic model obtained by identifying
\beq
z \sim e^{2i\pi/\e}z\, .
\eeq
It is equivalent to redefining the field $\vartheta \rightarrow \vartheta/\e$ while keeping its $2\pi$-periodicity. Taking this into account we need to dualize
\beq\label{fiber}
\Lag_{fiber}  = {r^2 \over 4\pi \e^2}(\partial_{\mu}\vartheta - \e b_{\mu})^2\, ,
\eeq
with $b_\mu$ treated as a background field. A detailed discussion of its T-dualization can be found in~\cite{Hori:2000kt}. Further references on the subject with direct connection to the model under consideration are~\cite{Alvarez:1993qi, Balog:1999ik}. The outcome is simply
\beq\label{Tdualfiber}
\Lag_{fiber} = {\e^2 \over r^2}\partial_{\mu}\varphi\partial^{\mu}\varphi +{\e \varphi \over \sqrt{\pi}}\,  \epsilon^{\mu \nu}\partial_{\mu}B_{\nu}\, ,
\eeq
now written in terms of the T-dual compact field $\varphi$. In the above expression we have made use of the following relation $\epsilon^{\mu \nu}\partial_{\mu}b_{\nu} = \epsilon^{\mu \nu}\partial_{\mu}B_{\nu}$ with $\epsilon^{\mu \nu}$ being the antisymmetric unit tensor. Note also that in our notations the field $\varphi$ has period $\sqrt{\pi}$, which is a convenient choice for the mapping with the bosonized fermion. The charge $\e$ could be eliminated by redefining the field $\varphi \rightarrow \varphi/\e$, hence changing the period of $\varphi$. It is associated to a discrete $\mathbb{Z}_\e$ symmetry which is spontaneously broken~\cite{Witten:1978bc} and is irrelevant in infinite volume. It leads however to interesting effects when the model is put on a finite cylinder, as we shall see in Section~\ref{BSM}. In view of this fact and in order to facilitate the mapping with the fermionic model we prefer to keep the charge $\e$ explicit as in~(\ref{Tdualfiber}).

As expected, the coordinate transformation leading from \eqref{fiber} to \eqref{Tdualfiber} is non-local. Explicitly it reads
\beq\label{SolPhi}
\varphi(\sigma, \tau) = {\sqrt{\pi} \over \e}\int_{\sigma}^{\infty}d\sigma' J_0(\sigma', \tau)\, ,
\eeq
where $J_\mu = r^2B_\mu/(2\pi) =  r^2(\e b_\mu-\partial_\mu \vartheta)/(2\pi \e)$ is the $U(1)$ current. An equivalent local form is given by
\beq
J_\mu = {\e \over \sqrt{\pi}}\epsilon_{\mu \nu}\partial^{\nu}\varphi\, .
\eeq
The $U(1)$ current is then automatically conserved in this picture and the associated charge is topological for the $\varphi$ field. A further interesting observation is that the T-dual bundle is trivial, as opposed to the original one. The topological structure of the original bundle translates however into interaction between $\varphi$ and the topological density $\epsilon^{\mu \nu}\partial_{\mu}B_{\nu} = i \epsilon^{\mu \nu}\partial_{\mu}\zb \partial_{\nu}z$ of the $CP^{\N-1}$ model.

Eventually, we can reinstate the dummy gauge field $A_\mu$ and write the T-dual of the model~(\ref{Lagrangian}) as
\beq\label{Tdual}
\Lag = {R^2 \over 4\pi}\D_{\mu}\bar{z}\D^{\mu}z + {R^2 -r^2 \over r^2R^2}\e^2\partial_{\mu}\varphi\partial^{\mu}\varphi +{\e\varphi \over \sqrt{\pi}}  \epsilon^{\mu \nu}\partial_{\mu}A_{\nu}\, ,
\eeq
where $D_\mu$ is the covariant derivative built with help of $A_{\mu}$. This is the form we will need to make contact with the bosonized version of the fermionic model. It is amusing to see how the $O(2\N)$ model is recovered in this picture. The latter corresponds to $r=R$ and at this point the field $\varphi$ in \eqref{Tdual} has no kinetic term. It becomes a Lagrange multiplier implementing
\beq
\epsilon^{\mu \nu}\partial_{\mu}A_{\nu} = 0\, .
\eeq
This in turn means that $A_\mu$ is pure gauge and can be eliminated by a local rotation of the $z$'s. Hence, locally, we recover the Lagrangian of the $O(2\N)$ model. A refined analysis would reveal that it is actually a model with target space $S^{2\N-1}/\mathbb{Z}_\e$. In the opposite limit $r \rightarrow 0$ the boson $\varphi$ decouples and we are left with the $CP^{\N-1}$ model.

\subsection{The fermionic model}\label{FermMod}

The fermionic extension of the $CP^{\N-1}$ model is defined by the Lagrangian~(\ref{fermionicmodels}). Apart from the inherited coupling $\kappa$, it contains two additional parameters: the fermion charge $\e$ and the Thirring coupling $\lambda$. They are both dimensionless. In this section we will relate them to the parameters of the bosonic theory, i.e., we will derive the equations \eqref{Intro:coup} quoted in the introduction. 

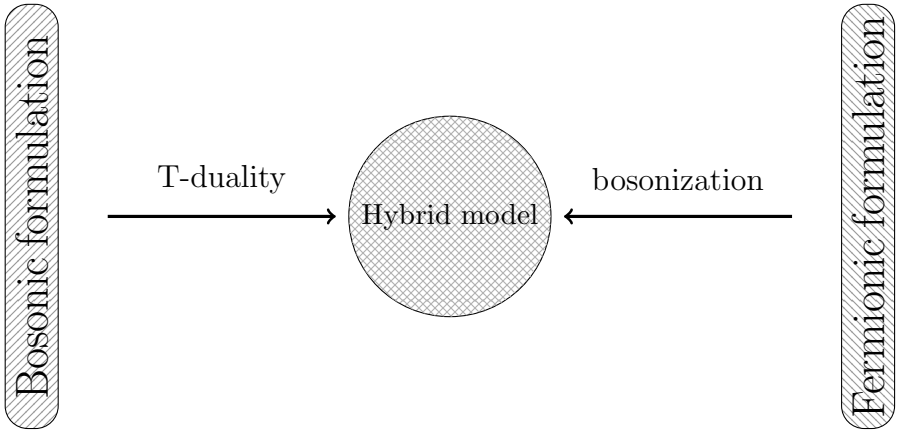
\begin{figure}
\begin{center}
\begin{tikzpicture}
%\textcolor{blue}{}
\draw[->,  very thick] (0,0) -- (3,0);
\draw[<-,  very thick] (6,0) -- (9,0);
\node at (1.5,0.5) {T-duality};
\node at (7.5,0.5) {bosonization};
\node at (-1,0) [rectangle, rotate = 90, draw=black,pattern = north east lines, pattern color = black!40, rounded corners=3mm,  text width=5.35cm] {\hspace*{2.5mm}\Large Bosonic formulation};
\node at (10,0) [rectangle, rotate = 90, draw=black, pattern = north west lines, pattern color = black!40, rounded corners=3mm, text width=5.35cm] {\Large Fermionic formulation};
\node at (4.5,0) [shape=circle, draw=black, pattern = north east lines, pattern color = black!30, postaction={pattern = north west lines, pattern color=black!30}]  {\small Hybrid model};
\end{tikzpicture}
\end{center}
\caption{The relation between the models.}
\label{BtoF}
\end{figure}

To make contact with the previous formulation we need to bosonize the fermion \cite{Coleman:1974bu}. The bosonized form of the model without Thirring coupling was already given in~\cite{Witten:1978bc}. Using the standard identity \cite{Coleman:1974bu}
\beq
\bar{\psi}\gamma_{\mu}\psi = {1\over \sqrt{\pi}}\epsilon_{\mu \nu}\partial^{\nu}\varphi \, ,
\eeq
where $\varphi$ stands for the bosonized fermion, one sees immediately that the net effect of the Thirring coupling is to redefine the kinetic term of the bosonized theory. Explicitly, the disguised fermionic Lagrangian takes the form
\beq
\mathcal{L}_{F} = {1+\lambda/\pi \over 2}\partial_{\mu}\varphi\partial^{\mu}\varphi + {\e \varphi \over \sqrt{\pi}} \epsilon^{\mu\nu}\partial_{\mu}A_{\nu}\, .
\eeq
It is identical to the T-dual version of the $\mathbb{Z}_\e$ quotient of the bosonic model \eqref{Tdual} if the couplings are identified as in~(\ref{Intro:coup}). The relation between the bosonic and fermionic models is summarized in Figure \ref{BtoF}.

The relation established above allows us to translate the physics of the bosonic model into the fermionic language. Firstly, the Thirring coupling $\lambda$, being essentially the same as the radius $r$ of the bosonic model, is not exactly marginal and runs with the renormalization scale. In the UV it approaches a constant value $\lambda_\infty$ which is a RG-invariant parameter of the model. Secondly, we learn that the \textit{true} deformation parameter of the fermionic model is neither $\e$ nor $\lambda_\infty$ on their own, but the combination
\beq\label{pferm}
\p = {\e^2 \over \N(1+\lambda_{\infty}/\pi)}\, .
\eeq
This quantity has to be identified with the parameter $\p=r^2_\infty/2\N$ introduced for the bosonic model. The relevance of this parameter for the fermionic model could be inferred from the large $\N$ analysis by generalizing the argumentation of \cite{Witten:1978bc} to non-zero value of the Thirring coupling. It is important to stress that the renormalization of this interaction is subleading at large $\N$ and may be ignored to leading order. One finds that $\p$ controls the mass of the gauge field, which is the only physical parameter in the theory. Varying the fermion charge $\e$ (assumed to be quantized) at a given value of $\p$ is equivalent to performing a $\mathbb{Z}_\e$-quotient of the bosonic model. As mentioned before, this operation has no effect on the model in infinite volume and, for instance, does not affect integrability.

This is the picture we inherit from the duality with the bosonic model. In what follows we will argue that one also recovers it directly from the fermionic formulation.

The fermionic model has naively more symmetries than the bosonic one. It has $U(1)\times SU(\N)$ symmetry from rotations of the $z$'s and $U(1)\times U(1)$ from vector$\times$axial rotations of the fermion. Due to the $U(1)$ gauge symmetry, one $U(1)$ subgroup is removed. This may be observed directly at the level of the Gauss law
\beq\label{BcFc}
2i\kappa \zb D_\mu z= \e \bar{\psi}\gamma_\mu \psi\, ,
\eeq
where $J_{\mu} = 2i\kappa \zb D_\mu z$ is the bosonic $U(1)$ current. There remains only one extra symmetry, which has no equivalent in the bosonic formulation. It derives from the axial $U(1)$ rotation of the fermion,
\beq\label{Axial}
\psi \rightarrow e^{i\alpha\gamma_5}\psi\, ,
\eeq
with $\gamma_5 = \gamma_0\gamma_1$. Fortunately, it is purely classical and is spoiled by the anomaly upon quantization. This is manifest in the bosonized form of the model, where the axial $U(1)$ transformation \eqref{Axial} translates to $\varphi \rightarrow \varphi + \alpha/\sqrt{\pi}$ and is not a symmetry of \eqref{Tdual}. What remains is a discrete chiral subgroup $\mathbb{Z}_{2\e}$ which is spontaneously broken down to $\mathbb{Z}_2$~\cite{Witten:1978bc}. They are generated by $\alpha = \pi/\e$ and $\alpha = \pi$, respectively. Up to the $\mathbb{Z}_2$ symmetry, the quantum symmetries of the fermionic model are then identical to those of the bosonic model quotiented by $\mathbb{Z}_{\e} \cong \mathbb{Z}_{2\e}/\mathbb{Z}_2$. We recall that discrete symmetries play no role in infinite volume. So before we put the theories on a finite cylinder in Section~\ref{BSM} the reader might simply ignore them.

The second point we wish to understand directly from the fermionic picture concerns the renormalization of the Thirring coupling $\lambda$. In the absence of interactions between bosons and fermion, i.e., when $\e=0$, the Thirring coupling is exactly marginal. We shall prove below  that this is not the case anymore if the interactions are switched back on. 

Before we do that we want to comment on the renormalization of the electric charge $\e$, or more precisely on the lack thereof. To make this point manifest in the fermionic picture one needs to take note of equation \eqref{BcFc}. This equation is an identity between the bosonic and fermionic $U(1)$ currents. They are physical observables and hence UV finite. As a consequence, the charge $\e$, which is just a proportionality factor between these two currents, should not renormalize.

Let us now turn to the Thirring coupling. To reveal its dependence on the renormalization scale we can consider the two-point function $\left<\bar{\psi}\gamma_\mu \psi(x)\bar{\psi}\gamma_\nu \psi(0)\right>$. The interesting property of this correlator is that it is sensitive to the Thirring coupling at lowest order in perturbation theory. It also follows from our previous discussion that it should be finite. This requirement is easily seen to be in conflict with a direct perturbative computation unless the Thirring coupling gets renormalized. This proves the assertion that $\lambda$ in~(\ref{fermionicmodels}) is a running coupling. The one-loop analysis is performed in Appendix \ref{RTC}. From it we deduce that the complete RG equations read
\beq\label{RGE}
\mu {\partial \kappa \over \partial \mu} = {\N \over 2\pi} + O(1/\kappa)\, , \qquad \mu{\partial \lambda \over \partial \mu} = {\e^2 \over 4\pi\kappa^2} + O(1/\kappa^3)\, .
\eeq
They agree with the bosonic ones~(\ref{RGEBM}) upon the identification~\eqref{Intro:coup}. The equation for the bosonic coupling $\kappa$ is the same as in the $CP^{\N-1}$ model. This will only be true at this order and  is actually a consequence of the fact that the correlator of fermionic currents is finite. Notice also that although the Thirring coupling does not appear explicitly in theses one-loop $\beta$-functions, no assumption was made about its actual value in the derivation. In other words, the equations \eqref{RGE} are valid to leading order in the $1/\kappa$ expansion, but for arbitrary value of  $\lambda$. We should nonetheless require $\lambda > -\pi$ for consistency \cite{Coleman:1974bu} with $\lambda \rightarrow -\pi$ corresponding to the $O(2\N)$ limit. Finally, we note that the solution to \eqref{RGE} is given by
\beq\label{RGEbis}
\kappa = {\N\over 2\pi}\log{(\mu/\Lambda)} + O(\log{\log{(\mu/\Lambda)}})\, , \qquad \lambda = \lambda_{\infty} - {\e^2\over 2\N \kappa} + O(1/\kappa^2)\, ,
\eeq
where the parameter $\lambda_{\infty}$ is the UV value of the Thirring coupling.
 
At the end it is quite natural to find that the Thirring coupling runs in this theory. The associated vertex is fully consistent with the symmetries of the theory. In fact, it is the only marginal interaction that we can add to a massless Dirac fermion minimally coupled to the $CP^{\N-1}$ model, up to an irrelevant $\theta$ term (see discussion in~\cite{Witten:1978bc}). One could also attempt to add mixed or bosonic current-current interactions. Due to the Gauss law~(\ref{BcFc}), they would not essentially differ from the Thirring interaction. The argument is somewhat heuristic but one can check explicitely that the two possible extra vertices are effectively redundant. The model \eqref{fermionicmodels} is hence the most general one with the given symmetries and thus the most likely to define a renormalizable theory.

\subsection{The case $\N=2$}\label{sec:PCF}

Before we discuss the integrability of the $U(1)\times SU(\N)$ model it is worthwhile to briefly summarize what happens at $\N=2$. This particular case provides a remarkable illustration of the previous generic considerations and has been extensively studied in the literature \cite{Wiegmann:1985jt, Cherednik, Balog:1999ik, Balog:2000wk, Forgacs:2000eu, Kawaguchi:2010jg, Kawaguchi:2011ub, Kawaguchi:2011pf, Kawaguchi:2012ve}.

To make contact with the standard formulation one just needs to repackage the two $SU(2)$ doublets into a single $SU(2)$ group element
\beq
\Omega =\left( \begin{array}{cc}
z_1 & -\bar{z}_2 \\
z_2 & \, \, \, \, \, \bar{z}_1 \end{array} \right)\, .
\eeq
Two natural sets of transformations on $\Omega$ are left and right $SU(2)$ rotations, $\Omega \rightarrow L \Omega R,$ with $L , R, \in SU(2)$. The left $SU(2)$ group rotates the multiplets $z_i$ and $\epsilon_{ij}\bar{z}_j$, while the right $SU(2)$ acts on $(z_1, -\bar{z}_2)$ and $(z_2, \bar{z}_1)$. The right transformations are not all symmetries of the model for generic value of $\eta$. This becomes manifest after rewriting the Lagrangian in terms of the right $SU(2)$ currents $j_\mu = \Omega^{-1}\partial_\mu \Omega$, 
\beq\label{PCF}
\Lag_{\N=2} = -{\kappa \over 2} \textrm{tr}\,  j_\mu j^\mu + {\kappa\eta \over 4}\left(\textrm{tr}\, j_{\mu}  \sigma_3\right)^2\,.
\eeq
In the above $\sigma_i$ denote Pauli matrices. One easily recognizes in \eqref{PCF} the deformation of the $SU(2)$ Principal Chiral Field (PCF) first considered by Wiegmann and Polyakov~\cite{Wiegmann:1985jt, Polyakov:1983tt}. For $\eta = 0$ the model is identical to the $SU(2)$ PCF model, well-known to be equivalent to the $O(4)$ sigma model.  Away from this point the second term in the Lagrangian breaks the right $SU(2)$ symmetry down to $U(1)$.

The model~(\ref{PCF}) is classically~\cite{Cherednik, Balog:2000wk, Kawaguchi:2010jg, Kawaguchi:2011ub, Kawaguchi:2011pf, Kawaguchi:2012ve} and quantum-mechanically~\cite{Wiegmann:1985jt, Balog:1999ik, Balog:2000wk, Forgacs:2000eu} integrable for any value of the deformation parameter $\p$. Its S-matrix was derived  by Wiegmann~\cite{Wiegmann:1985jt} and found to be given by the tensor product
\beq\label{Sn2}
\mathbb{S} = \mathbb{S}_{SU(2)} \otimes \mathbb{S}_{SG}^{(\p)}\, ,
\eeq
where the first factor is the minimal $SU(2)$ S-matrix and the second one coincides with the sine-Gordon S-matrix with the parameter $\p$.%
\footnote{We refer the reader to \cite{Zamolodchikov:1994uw} for a definition of this parameter in the context of the sine-Gordon theory.} The value $\p=1$ corresponds to the free Dirac point for the sine-Gordon S-matrix. At this point the S-matrix \eqref{Sn2} simplifies considerably and becomes the minimal reflectionless $U(2)$ S-matrix. For $p<1$ the theory has spinon-antispinon bound states~\cite{Wiegmann:1985jt, Balog:1999ik}. These breathers come in pair (singlet, vector) w.r.t. $SU(2)$ and are degenerate in mass. The lightest breathers consist of a $C$-odd singlet, which can be seen as the dual field $\varphi$ or equivalently $\epsilon_{\mu \nu}\partial^{\mu}\bar{z}\partial^{\nu}z$, and a $C$-even vector created by $X_i \sim \bar{z}\sigma_i z$. At small enough $\p$ these bound states are the lightest excitations in the spectrum. At $\p=0$ they decouple from one another, with the singlet becoming a free massive field and the vector describing the $O(3)\simeq CP^{1}$ sigma model. Up to the accidental degeneracy, this picture is remarkably close to what is expected to happen for higher values of $\N$. In the latter case, however, integrability will place severe restrictions on the deformation parameter $\p$, as will argue in the following section.

Surprisingly, the S-matrix~(\ref{Sn2}) cannot be found in the $U(\N)$ classification of Berg et al.~\cite{Berg:1977dp}. A natural question is whether this classification could be missing a one-parameter family of $U(\N)$ solutions. Such a family would be particularly significant for the class of models we are considering. We clarify this point in Appendix~\ref{Smatrix}, where it is shown that the Wiegmann's solution does not admit a lift beyond $\N=2$ in agreement with the conclusion of Berg et al.~\cite{Berg:1977dp}. We nevertheless found a caveat in their classification for $\N=2$, which allows us to embed the S-matrix~(\ref{Sn2}) into the space of $\N=2$ solutions.

It is also worthwhile considering the fermionic formulation of the $\N=2$ model at the special point $\p=2$. To understand what is special about this value, we recall \cite{Kawaguchi:2011ub} that the complete symmetry of the model is $SU(2)\times SU_{q}(2)$, where $SU_q(2)$ is the quantum deformation of $U(\mathfrak{su}_2)$ with $q = \exp{(i\pi/\p)}$. This symmetry is somewhat visible in the decomposition \eqref{Sn2} if one recalls that $SU_{q}(2)$ is the quantum symmetry group of the sine-Gordon theory \cite{Bernard:1990ys}. It is however not manifest at the level of the Lagrangian and for  $q\neq 1$ it is implemented by non-local currents \cite{Kawaguchi:2011ub}, see also \cite{Bernard:1990ys}. At $\p=2$ the $SU_q(2)$ symmetry becomes isomorphic to the centrally extended $\mathcal{N}=2$ SUSY algebra with $4$ supercharges. Interestingly enough, this SUSY algebra can be realized linearly in the fermionic model. This requires $\e = 2$, when the theory has two conserved spin $3/2$ supercurrents proportional to $\epsilon_{ij}z^{i}D_\nu z^{j}\gamma^\nu \gamma_\mu \psi $ and its complex conjugate. What is this model? It turns out that this is the SUSY $CP^1$ model. This is not the standard formulation of this theory, which by definition is equipped with a $SU(2)$ doublet of Dirac fermions constrained by $\bar{z}_i \chi_i = z_j \bar{\chi}_j = 0$, nevertheless it is equivalent \cite{Witten:1977xn}. The key observation is that one can solve these constraints in terms of a single Dirac field $\psi$ defined such that
\beq
\bar{\chi}_i = \psi \epsilon_{ij}z_j\, , \qquad \chi_i = \bar{\psi} \epsilon_{ij}\bar{z}_j\, .
\eeq
This fermion has charge $\e=2$ in our notations. The supercurrents mentioned above are then nothing else than the supercurrents of the SUSY $CP^1$ model expressed in terms of the fermion $\psi$. Interestingly, the Thirring coupling for this model is fixed by supersymmetry
\beq
\lambda = -{1\over \kappa}\, .
\eeq
As a consistency check of our previous results, we observe that it is in perfect agreement with~(\ref{pferm}) and~(\ref{RGE}) if $\N=\p=\e=2$. This concludes our discussion of the case $\N=2$.

\section{Integrability}\label{Int}

In this section we study integrability of the fermionic and the bosonic model for generic $\N$. We will find that it is a sporadic phenomenon requiring fixing the parameter $\p$. 

\subsection{Classical integrability}

We will start with the classical integrability. It turns out that only the fermionic model is classically integrable for generic value of the deformation parameter. This may appear contradictory at first, as we have just shown the equivalence of the both models. We will resolve this paradox in Subsection \ref{sec:fermform}.

\subsubsection{Bosonic model}\label{IBM}

Since the actual analysis is slightly technical, we will first offer a summary of the results. Customarily, classical integrability of a physical system relies on the existence of a non-abelian current $\tilde{j}^{\phantom{.} ij}_\mu$, which is both \textit{flat} and \textit{conserved},
\beq\label{flateq}
\partial_\mu \tilde{j}^{ij}_{\nu}-\partial_\nu \tilde{j}^{ij}_{\mu}+\left[\tilde{j}_{\mu},\tilde{j}_{\nu}\right]^{ij}=0\, , \qquad \partial^{\mu}\tilde{j}^{\phantom{.} ij}_\mu = 0\, .
\eeq
Its existence allows one to construct the Lax connection
\beq \label{Lbyj}
L_{\mu}(x) = {1\over 1-x^2}\, \tilde{j}_{\mu} + {x\over 1-x^2} \, \epsilon_{\mu \nu} \tilde{j}^{\nu} \,,
\eeq
which is automatically flat for any value of the spectral parameter $x$. The latter property guarantees the conservation of infinitely many non-abelian charges, which are manifestation of the integrability of the model. They may be generated with help of the monodromy matrix defined as the path-ordered exponential of the Lax connection
\beq \label{monodromy}
M(x) = P\,{\textrm{exp}}{\int d\sigma L_\sigma (x)}\, .
\eeq
As will be shown below, for the bosonic model the current $\tilde{j}_{\mu}$ may only be constructed in three exceptional cases. These special cases correspond to fixing \textit{one} of the two parameters of the classical theory: $\eta$ and $N$. The case of $\eta=0$ is perhaps the most obvious one since the Lagrangian \eqref{Lagrangian} reduces to that of the $O(2\N)$ sigma model, which is well known to be classically integrable. Due to the symmetry enhancement this case is actually not covered by our analysis for generic $\N$. The value $\eta = 1$ corresponds to the $CP^{\N-1}$ model and is thus also an integrable case. Our analysis below shows that as long as $N$ is arbitrary, there are no other special values of $\eta$. The situation is very different if and only if $N=2$. In this case the classical integrability is present irrespectively of the value of the parameter $\eta$. This is due to a special kinematical relation which ceases to hold for higher values of $N$. 

To substantiate the above picture, we will construct the most general conserved current and impose the flatness condition. For an easier comparison with the fermionic theory, it is convenient to work with the formulation \eqref{Lagrangian} of the bosonic model and to introduce the couplings $\eta$ and $\kappa$ defined previously. The equations of motion are then identical to the ones describing the $CP^{\N-1}$ model
\beq\label{eom}
D_\mu D^{\mu}z^{i} +D_{\mu}\bar{z}D^{\mu}z\, z^{i} = 0\, .
\eeq
The sole difference is that the gauge field
\beq
A_{\mu} = i\eta \bar{z}\partial_{\mu}z
\eeq
now depends on the deformation parameter $\eta$. Since we are discussing the classical theory, we can further set $\kappa=1$. The currents corresponding to the global $U(\N)$ symmetry of the model can then be written as
\beq
j^{\phantom{.} ij}_\mu= \zb^i D_\mu z^j -  D_{\mu}\zb^i z^j \, .
\eeq
Notice that these currents are not real because we rescaled them by a factor $-i$. It is also convenient to single out the $U(1)$ component
\beq
j'_\mu= \Tr\,  j_{\mu} = 2\zb D_{\mu} z\, ,
\eeq
which was previously denoted $-iJ_{\mu}$ in Section \ref{TM}. An important remark regarding this current is that it is \textit{not} axially conserved
\beq\label{AxialA}
\epsilon^{\mu\nu}\partial_{\mu} j'_\nu = 2(1-\eta)\epsilon^{\mu\nu}D_\mu\zb D_\nu z = 2(1-\eta)\epsilon^{\mu\nu}\partial_\mu\zb \partial_\nu z\, .
\eeq
An exceptional case is $\eta =1$ when the $U(1)$ current is actually zero.

The most general conserved current is a combination of the above currents and a topological term~\cite{Balog:2000wk, Kawaguchi:2010jg}
\beq\label{gcurrent}
\tilde{j}^{\phantom{.} ij}_{\mu}= a\, j^{\phantom{.} ij}_{\mu}+b\, j'_{\mu}\delta^{ij}+ \epsilon_{\mu\nu}\partial^{\nu} f^{ij}\,,
\eeq
where $a, b$ are both constant and $f$ is a matrix function.  For $\eta = 0$ the symmetry is enhanced to $O(2\N)$ and hence the theory has extra conserved currents. We shall not consider this case in what follows. It is important to point out that for $\eta=1$ the parameter $b$ is irrelevant since $j'_{\mu} = 0$. Finally, the function $f$ in \eqref{gcurrent} should have mass dimension zero. Its most general form is then given by
\beq
f^{ij} =f(\zb z) \zb^i z^j = c\,\zb^i z^j\, ,
\eeq
where $c = f(\zb z) =f(1)$ is arbitrary. If one can choose $a, b, c$ in such a way that the current~(\ref{gcurrent}) is flat, it will immediately allow us to construct the Lax connection~(\ref{Lbyj}).

Let us check whether such a choice is possible. Using \eqref{eom}, the current \eqref{gcurrent} is easily found to obey
\beq\label{flatf}
\partial_\mu \tilde{j}^{ij}_{\nu}-\partial_\nu \tilde{j}^{ij}_{\mu}+\left[\tilde{j}_{\mu},\tilde{j}_{\nu}\right]^{ij}=\epsilon_{\mu\nu}\sum_{k=1}^{4}\alpha_k O^{ij}_{k} - \epsilon_{\mu \nu}(1-a)\Box f^{ij} \, .
\eeq
Here the operators $O^{ij}_{k}$ are defined by
\beq\label{Op2}
\begin{aligned}
O_1^{ij} &= \epsilon^{\rho\sigma}D_\rho\zb^{i}D_{\sigma}z^j\, , \qquad \, \, \, \, \, \, \, \, \, \, O_2^{ij} = \epsilon^{\rho\sigma}D_\rho \bar{z}D_{\sigma}z\, \zb^{i}z^{j}\, ,\\
O_3^{ij} &= \epsilon^{\rho\sigma}\bar{z}D_{\rho}z\, \partial_{\sigma}(\zb^{i}z^{j})\, , \qquad O_4^{ij} = \epsilon^{\rho\sigma}D_{\rho}\zb D_{\sigma} z\ \delta^{ij}\, ,\\
\end{aligned}
\eeq
with $O^{ij}_3$ being identically zero if $\eta=1$. The coefficients $\alpha_k$ in \eqref{flatf} are given by
\beq\label{alpha}
\begin{aligned}
&\alpha_1 = c^2-a(2-a)\, , \qquad \alpha_2 = a(2\eta-a)-c^2\, , \\
&\alpha_3 = c^2-a^2\, , \qquad \qquad \, \, \, \, \alpha_4 = (\eta-1)b\, .
\end{aligned}
\eeq
If the current $\tilde{j}_{\mu}$ is flat, then the last term in the r.h.s of~(\ref{flatf}) should vanish separately, as it is the only term which is even under the charge conjugation $z\leftrightarrow \zb$. This leads to two possibilities: $a=1$ or $c=0$. Imposing now that the sum in \eqref{flatf} vanishes term by term yields two solutions to the flatness equation~(\ref{flateq}):
\beq\label{flatcond}
\begin{aligned}
&a=1\, , \qquad c^2=1\, , \qquad \eta=1\,, \quad \textrm{or}\\
&a=2\, , \qquad c=0\, , \qquad \, \, \,  \eta=1\, .
\end{aligned}
\eeq
These two solutions produce the same charges after expanding the monodromy matrix~(\ref{monodromy}) at large $x$ and are consequently equivalent. The solution with $a=2$ is the conventional expression for the Lax connection of the $CP^{\N-1}$ sigma model.

As it is clear from \eqref{flatcond} no solution is found for $\eta \neq 1$. This disappointing conclusion can be avoided if and only if the operators~(\ref{Op2}) are linearly dependent. In fact, the above-listed operators form a basis of dimension two operators that are antisymmetric with respect to charge conjugation. Any reductiveness of this basis would imply
\beq
\sum^{4}_{n=1} c_n\,O^{ij}_n = 0\,,
\eeq
for some values of $c_n, \, n=1,2,3,4$. The complete set of constraints on the coefficients $c_n$ may be derived by contracting with $\delta^{ij}, \zb^{j} z^{i}, \zb^{j}D_\alpha z^{i}$ and $D_\alpha \zb^{j} D^{\alpha} z^{i},$ respectively. One finds the following set of equations
\beq\label{oprel}
(\N-2)\, c_2=0\, ,\qquad  \left(O_1 + O_2+ O_3 - O_4\right)c_2 =0\,.
\eeq
Clearly, if any, the non-trivial solution exists only for $\N=2$. An explicit parameterization in this case allows to verify that \eqref{oprel} is satisfied for any $c_2$. The current \eqref{gcurrent} is then found to be flat for any value of $\eta$ if $a=1, c^2 = \eta$, and $b=-1$. The two solutions $c=\pm \eta$ are related by charge conjugation and generate the same charges. This flat and conserved current, albeit in a different formulation, has been found in \cite{Forgacs:2000eu} and independently in \cite{Kawaguchi:2011ub}.

The above construction suggests that the classical integrability is not present when $\eta \neq 1$ and $\N > 2$. Before discussing the fermionic formulation let us delve into the possibility of having more general Lax connection than~\eqref{Lbyj}. We are thus led to study the most general solution to the flatness condition
\beq \label{flatL}
\partial_{\mu} L_{\nu} - \partial_{\nu} L_{\mu} + [L_{\mu}, L_{\nu}]=0\,,
\eeq
modulo the kinematic constraint $\zb z=1$ and the equations of motion \eqref{eom}. For $\eta \neq 1$ and $\N>2$ we found that no non-trivial solutions exist, up to the gauge transformation generated by the local unitary matrix $U^{ij} = \delta^{ij}+\alpha \zb^{i}z^j$. This corroborates the claim that there is no classical integrability for generic values of $\eta$ and $\N$. 

\subsubsection{Fermionic model}\label{sec:fermform}

The situation looks differently if one chooses \eqref{fermionicmodels} instead of \eqref{Lagrangian} as the starting point. We have seen that the $CP^{\N-1}$ model is integrable. The Lagrangian \eqref{fermionicmodels} is an extension of that model with a massless Dirac fermion and it is known that one can construct a flat and conserved current in this case~\cite{Abdalla:1981yt}. Let us briefly present this construction. The equations of motion for the fields $z^{i}$ are the same as before~(\ref{eom}) and so is the expression for the $U(\N)$ currents. The gauge field now also accounts for the fermion
\beq
A_\mu = i\zb \partial_\mu z -{\e \over 2\kappa}\bar{\psi}\gamma_\mu \psi\, .
\eeq
Note that the bosonic $U(1)$ current, being proportional to $\bar{\psi}\gamma_\mu \psi$, is \textit{axially} conserved at the classical level
\beq\label{ACC}
\epsilon^{\mu \nu}\partial_{\mu}j'_{\nu} = 0\,.
\eeq
This makes the analysis very similar to the $CP^{\N-1}$ model, apart from the fact that the $U(1)$ current does not have to vanish in the fermionic model. Our previous analysis carries over to this case, but the coefficients $\alpha_k$ appearing in \eqref{flatf} now need to be replaced by their values \eqref{alpha} evaluated at $\eta=1$. The immediate conclusion is that the conserved current~(\ref{gcurrent}) is flat for $c^2=a=1$ and for any value of $b$. The latter parameter controls the $U(1)$ part of the mondromy matrix which is not of interest here. Fixing $b=0$ and choosing any one of the two values $c=\pm 1$ one easily generates the non-abelian charges constucted in~\cite{Abdalla:1981yt}, by expanding the mondromy matrix~(\ref{monodromy}) at large $x$. Notice that the above construction is independent of the value of the Thirring coupling, since it only entails the axial conservation law~(\ref{ACC}).

As already pointed out at the beginning of this discussion, the classical integrability of the fermionic model does not contradict the non-existence of the corresponding current for the bosonic formulation \eqref{Lagrangian}. The reason is that the bosonization is \textit{not} an innocent classical transformation: it automatically accounts for the axial anomaly, which in turn spoils the classical integrability of the bosonic model.

The above observation has far-reaching consequences. If the classical bosonic theory incorporates the effect of the axial anomaly of the fermionic model, should we not conclude that its description is more reliable and that integrability can only occur at $\eta=1$? We believe this not the case. The reason is that the classical bosonic theory only resums a subclass of quantum corrections of the fermionic model. To make this point more precise we observe that the classical fermionic and bosonic descriptions probe different parts of the parameter space. This becomes clear after recalling the relation between the couplings of both theories,
\beq\label{etap}
\eta = 1- {\N\p \over 2\pi \kappa} + \ldots\, ,
\eeq
where dots stand for quantum corrections $\sim 1/\kappa^2$. The fermionic model becomes classical for $\kappa \gg 1$. In this limit the parameter $\eta$ takes the \textit{classical} value $\eta =1$ for any finite value of $\p$. Moving away from this value requires $\p \sim \kappa$. Since we found that classical integrability is not a property of the bosonic model for $\N >2$, it is reasonable to infer that the fermionic theory is not integrable at the quantum level for arbitrarily large values of $\p$. Otherwise the phenomenon should be visible in the corresponding semiclassical regime of the bosonic theory. A remarkable illustration is the case $\N=2$, for which the theory is quantum integrable for any $\p$ and classically integrable for any $\eta$.

\subsection{Quantum integrability}

We have found that classical integrability of the fermionic model does not imply classical integrability for its bosonic dual. Moreover, it seems unlikely that the fermionic model will remain quantum integrable at large values of $\p$. In this section we will argue that this conclusion should also apply for more generic values of $\p$, hence leaving little room for integrability when $\N>2$. The large $\N$ analysis will reveal that integrability is only achieved at $\p=1$ and $\p=\infty$. 

\subsubsection{Counting and fine tuning} \label{sec:first glimpse}

An elegant way of providing evidence for the quantum integrability of a certain class of models offers the counting argument  advocated in~\cite{Goldschmidt:1980wq}. The underlying idea is the following. Observe that the classical theory is scale-invariant. Then in light-cone coordinates the conservation of the stress-energy tensor takes the form $\pa_{+} T_{--}=0$. This for example trivially implies
\beq \label{Tmmbis}
\pa_{+} \left(T_{--}\right)^n=0\,,
\eeq
for any positive integer $n$. If there exist some axially conserved currents in the theory they may also be taken into account in the product. The scale invariance of \eqref{fermionicmodels} is broken at the quantum level. But for integrable models the above conservation law is expected to become deformed not spoiled. Indeed, as long as the deformation is of the form
\beq \label{dTmmbis}
\pa_{+} \left((T_{--})^n + F_1\right)=\pa_{-} F_2\,,
\eeq
for some operators $F_1$ and $F_2$, it leads to a conserved quantum charge. It was proposed in~\cite{Goldschmidt:1980wq} to compose a list of all quantum anomalies modulo kinematical constraints and equations of motion that can appear on the r.h.s of \eqref{Tmmbis}. The next step is to list all admissible operators of the form $\pa_{+} (...)$ and $\pa_{-} (...)$. If the both lists match, every anomaly may be expressed as a divergence of some operator, consequently allowing to write \eqref{dTmmbis} at the quantum level.

We have classified all possible anomalies and divergences for two of the higher conservation laws and found that in both cases there is \textit{one} unmatched anomaly if $\N>2$. We present the lists for one of the higher conservation laws in Appendix~\ref{sec:CA}. In principle, there might exist a higher charge other than the ones considered by us for which the counting would go through. However, in light of the fact that one of the conservation laws we have studied was sufficient to argue for integrability for variety of models \cite{Goldschmidt:1980wq} we consider this as a strong evidence against the integrability of \eqref{fermionicmodels} for generic values of $\N$ and $\p$.

We would like to stress that the counting \textit{does not} depend on the values of the parameter $\p$. Since the lists differ only by one entry and the theory has one free parameter $\p$, it may happen that the coefficient of the unmatched anomaly cancels out for some adjusted values of $\p$. We suspect that this mechanism will restore the integrability for $\p=1$ and shall give evidence for it in the following. It should also be observed that at $\p=\infty$, when the symmetry is enhanced to $O(2\N)$, the counting argument leads to correct conclusions  \cite{Goldschmidt:1980wq}. Notice finally that the case $\N=2$ appears again to be exceptional. At this value one needs to impose an additional kinematical relation \cite{Goldschmidt:1980wq}, which reduces by one the number of anomalies in the list. The higher conservation law is then guaranteed to exist at the quantum level, independently of the value of $\p$.

\subsubsection{Large $\N$ analysis}

To shed light on the possible values of the parameter $\p$ for which the model becomes integrable, we shall look at the spinon-(anti)spinon elastic scattering amplitudes at large $\N$. Whether the theory is integrable or not, the two-to-two scattering processes are always constrained by the kinematics in two dimensions. The two incoming momenta ought to be conserved separately and the amplitudes are functions of a single Mandelstam invariant. In the case of the $U(1)\times SU(\N)$ model we have three distinct elastic processes to consider. They are associated with the S-matrix elements
\beq
\begin{aligned}
\left< p_{3k}, p_{4l}|\mathbb{S}|p_{1i}, p_{2j}\right> &= S_{ij}^{kl}(\theta) \delta(p_1-p_3)\delta(p_2-p_4) + S_{ij}^{lk}(\theta)  \delta(p_1-p_4)\delta(p_2-p_3) \, ,\\
\left< p_{3k}, \bar{p}_{4l}|\mathbb{S}|p_{1i}, \bar{p}_{2j}\right> &= F_{ij}^{kl}(\theta) \delta(p_1-p_3)\delta(\bar{p}_2-\bar{p}_4) + B_{ij}^{lk}(\theta)\delta(p_1-\bar{p}_4)\delta(\bar{p}_2-p_3)\, ,
\end{aligned}
\eeq
where the asymptotic states are normalized as $\left< p'|p\right> = \delta(p-p')$. Here $p, \bar{p}$ are spinon, antispinon momenta and $\theta = 2\, \textrm{arcosh}(s/4m^2)$ is related to the square of the center-of-mass energy $s$. 

The symmetry of the problem permits to decompose the matrices $S, F$ and $B$ into $6$ scalar amplitudes~\cite{Berg:1977dp}
\beq\label{SUnDec}
\begin{aligned}
S_{ij}^{kl}(\theta) &= u_1(\theta) \delta_{i}^{k}\delta_{j}^{l} + u_2(\theta) \delta_{i}^{l}\delta_{j}^{k}\, , \\ 
F_{ij}^{kl}(\theta) &= t_1(\theta) \delta_{i}^{k}\delta_{j}^{l} \, + t_2(\theta) \delta_{ij}\delta^{kl}\, , \\
B_{ij}^{lk}(\theta) &= r_1(\theta) \delta_{i}^{k}\delta_{j}^{l} \, + r_2(\theta) \delta_{ij}\delta^{kl}\, . 
\end{aligned}
\eeq
They are not all independent due to the crossing symmetry between $s$ and $t = 4m^2-s$ channel
\beq\label{Crossing}
u_{1,2}(i\pi-\theta) = t_{1,2}(\theta)\, , \qquad r_1(i\pi-\theta) = r_2 (\theta)\, .
\eeq

\begin{figure}
\begin{center}
\begin{tabular}{x{5cm}x{5cm}x{5cm}}
\begin{tikzpicture}[scale=1]
\pgfsetlinewidth{\wid}
\draw [decoration={markings, mark=at position 0.4 with {\arrow{stealth}}},
        postaction={decorate} ] (0,1) -- (2,1);
\draw [decoration={markings, mark=at position 0.4 with {\arrow{stealth}}},
        postaction={decorate} ] (0,-1)--(2,-1);
 \filldraw (2,1) circle (0.5mm);
 \filldraw(2,-1) circle (0.5mm);
 \draw [decorate, decoration = {zigzag,amplitude=1mm}] (2,1) -- (2,-1);
 \draw[decoration={markings, mark=at position 0.75 with {\arrow{stealth}}},
        postaction={decorate} ] (2,1)--(4,1);
\draw [decoration={markings, mark=at position 0.75 with {\arrow{stealth}}},
        postaction={decorate} ] (2,-1) -- (4,-1);
\node[font = \fontsize{8}{8}] at (0.5,1.2) {$p_{1,i}$}; 
\node[font = \fontsize{8}{8}] at (0.5,-1.2) {$p_{2,j}$};         
\node[font = \fontsize{8}{8}] at (3.5,1.2) {$p_{1,k}$}; 
\node[font = \fontsize{8}{8}] at (3.5,-1.2) {$p_{2,l}$};
\end{tikzpicture} 
&

\begin{tikzpicture}[scale=1]
\pgfsetlinewidth{\wid}
\draw [decoration={markings, mark=at position 0.4 with {\arrow{stealth}}},
        postaction={decorate} ] (0,1) -- (2,1);
\draw [decoration={markings, mark=at position 0.4 with {\arrow{stealth}}},
        postaction={decorate} ] (0,-1)--(2,-1);
 \filldraw (2,1) circle (0.5mm);
 \filldraw(2,-1) circle (0.5mm);
 \draw [decorate, decoration = {zigzag,amplitude=1mm}] (2,1) -- (2,-1);
 \draw[decoration={markings, mark=at position 0.75 with {\arrow{stealth}}},
        postaction={decorate} ] (2,1)--(4,1);
\draw [decoration={markings, mark=at position 0.75 with {\arrow{stealth}}},
        postaction={decorate} ] (2,-1) -- (4,-1);
\node[font = \fontsize{8}{8}] at (0.5,1.2) {$p_{1,i}$}; 
\node[font = \fontsize{8}{8}] at (0.5,-1.2) {$p_{2,j}$};         
\node[font = \fontsize{8}{8}] at (3.5,1.2) {$p_{2,l}$}; 
\node[font = \fontsize{8}{8}] at (3.5,-1.2) {$p_{1,k}$};
\end{tikzpicture} 
&\begin{tikzpicture}[scale=1]
\pgfsetlinewidth{\wid}
\draw [decoration={markings, mark=at position 0.4 with {\arrow{stealth}}},
        postaction={decorate} ] (0,1) -- (2,1);
\draw [decoration={markings, mark=at position 0.4 with {\arrow{stealth reversed}}},
        postaction={decorate} ] (0,-1)--(2,-1);
 \filldraw (2,1) circle (0.5mm);
 \filldraw(2,-1) circle (0.5mm);
 \draw [decorate, decoration = {zigzag,amplitude=1mm}] (2,1) -- (2,-1);
 \draw[decoration={markings, mark=at position 0.75 with {\arrow{stealth}}},
        postaction={decorate} ] (2,1)--(4,1);
\draw [decoration={markings, mark=at position 0.75 with {\arrow{stealth reversed}}},
        postaction={decorate} ] (2,-1) -- (4,-1);
\node[font = \fontsize{8}{8}] at (0.5,1.2) {$p_{1,i}$}; 
\node[font = \fontsize{8}{8}] at (0.5,-1.2) {$p_{2,j}$};         
\node[font = \fontsize{8}{8}] at (3.5,1.2) {$p_{2,k}$}; 
\node[font = \fontsize{8}{8}] at (3.5,-1.2) {$p_{1,l}$};
\end{tikzpicture} 
\end{tabular}
\caption{\label{TreeLevel} Tree level Feynman diagrams for the scattering amplitudes $u_1(\theta), u_{2}(\theta)$ and $r_{1}(\theta)$, respectively. Zigzag lines stand for both scalar and vector exchange.}
\end{center}
\end{figure}
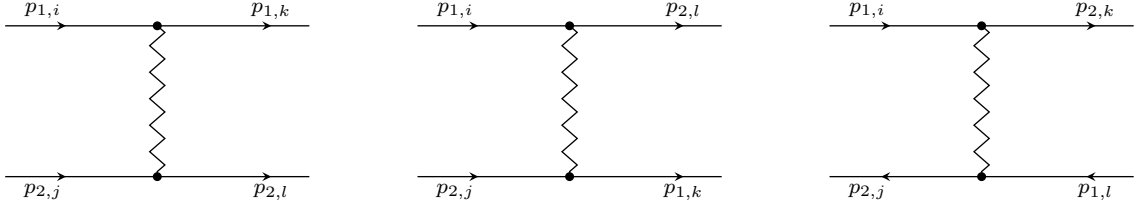

At large $\N$ these amplitudes read
\beq\label{FreeLimit}
u_1(\theta) = t_1(\theta) = 1\, , \qquad u_{2}(\theta) = t_{2}(\theta) = r_1(\theta) = r_2 (\theta) = 0\, ,
\eeq
as in any weakly coupled (free) theory. The computation of the leading $1/\N$ corrections is a direct application of the Feynman rules, which are the same as for the $CP^{\N-1}$ model~\cite{D'Adda:1978uc} but with the gauge-field propagator replaced by~(\ref{gpropagator}). Due to the crossing symmetry~(\ref{Crossing}) we can further restrict ourselves to the evaluation of $u_{1,2}(\theta)$ and $r_{1}(\theta)$. The relevant Feynman diagrams are depicted in Fig.~\ref{TreeLevel}.
One should pay attention to the fact that conventional Feynman algebra computes the amplitude $\mathcal{M}_{ij}^{kl}$ entering%
\footnote{We recall that we are working with the normalization $\left< p'|p\right> = \delta(p-p')$.}
\beq
\left< p'_{1k}, p'_{2l}|\mathbb{S}-1|p_{1i}, p_{2j}\right> = {1\over 4E_1E_2}\delta(E-E')\delta(p-p') \mathcal{M}_{ij}^{kl}(\theta)\, ,
\eeq
where $E$ and $p$ are total energy and momentum, respectively. Hence when relating the Feynman amplitude $\mathcal{M}_{ij}^{kl}$ to the one introduced before, the Jacobian associated to the change of measure has to be taken into account. It leads to
\beq
S_{ij}^{kl}(\theta) = \delta_{i}^{k}\delta_{j}^{l}+ {\mathcal{M}_{ij}^{kl}(\theta) \over 2\sqrt{-st}} \, ,
\eeq
and similar expressions for the spinon-antispinon backward and forward S-matrix: $B_{ij}^{kl}(\theta)$ and $F_{ij}^{lk}(\theta)$.  Summing up the diagrams in Figure \ref{TreeLevel} gives
\beq
\begin{aligned}\label{ScattAmp}
u_1(\theta) &= 1 - {i\pi \over \N}\left({2m^2 \over \sqrt{-st}} + {s-2m^2 \over \p \sqrt{-st}}\right)+O\left(\frac{1}{N^2}\right)\, , \\
u_2(\theta) &= \, \, \, \, -\, {i\pi \over \N}\left({1 \over \theta} + {s \over \theta s+2(\p-1)\sqrt{-st}}\right)\, +O\left(\frac{1}{N^2}\right), \\
r_1(\theta) &= \, \, \, \, -\, {i\pi \over \N}\left({1 \over \theta} - {s \over \theta s+2(\p-1)\sqrt{-st}}\right)+O\left(\frac{1}{N^2}\right)\, .
\end{aligned}
\eeq
In each of these equations, the first/second term in brackets originates from the scalar/vector exchange. In the limit $\p \rightarrow \infty$ the vector exchange is suppressed and we are left with the $O(2\N)$ sigma model results written in $SU(\N)$ variables. In the opposite limit, when $\p\rightarrow 0$, the spinon-spinon scattering amplitude $u_1(\theta)$ diverges. This phenomenon is a manifestation of the charge confinement in the  $CP^{\N-1}$ sigma model.

For generic value of $\p$ the scattering amplitudes \eqref{ScattAmp} are not consistent with factorized scattering, which is the salient property of integrable theories. It can be shown \cite{Berg:1977dp} indeed that the factorizability of the spinon-spinon scattering leads to
\beq\label{FSC}
u_{2}(\theta) = -{i\nu \over \theta}u_{1}(\theta)\,,
\eeq
for some constant $\nu$. The origin of this relation is recalled in Appendix \ref{Smatrix} for completeness. There are only two possible values of $\p$ at which the scattering amplitudes \eqref{ScattAmp} obey the relation~(\ref{FSC}): $\p=\infty$ and $\p=1$. Leaving aside the point $\p=\infty$, we conclude that integrability at large $\N$ requires $\p=1$. The analysis at this point becomes identical to the one carried out by K\"oberle and Kurak \cite{Koberle:1987wc}. The large $\N$ amplitudes \eqref{ScattAmp} are then found to agree with the large $\N$ expansion of the minimal reflectionless S-matrix of \cite{Berg:1977dp}, for which $r_1(\theta) = r_2(\theta) =0$.

Our analysis above was focused on the scattering amplitudes of spinons, because they are the only stable asymptotic excitations for $p\geqslant 1$. As far as the integrability is concerned it does not really matter whether additional excitations can form in the complementary domain $\p < 1$. From a broader perspective it is nonetheless interesting to see what we can learn about the spectrum in this domain by considering the scattering amplitudes~(\ref{ScattAmp}). We already know that for $\p < 1$ another stable excitation is carried by the gauge field $A_\mu$. It describes a $C$-odd, $SU(\N)$ singlet whose mass is below the two-spinon threshold when $p < 1$. At small enough $p$ this boson is actually the lightest excitation in the spectrum since then $M^2 \simeq 12pm^2$~\cite{Witten:1978bc}. It is the large $\N$ relative of the singlet bound state found at $\N=2$, despite the fact that the mass of the latter exhibits a rather different scaling $M^2 \sim p^2m^2$ at small $p$~\cite{Balog:1999ik}. This feature seems to be tied to peculiarities of the instantons gas for $\N=2$. Other spinon-antispinon bound states will form for $p<1$, though, as opposed to $\N=2$, none of them becomes degenerate with the gauge field. An analysis of the pattern of these bound states based on an effective Schr\"odinger equation may be found in \cite{Campostrini:1993fr}. We can have a glimpse at one of them by mapping the non-relativistic limit $\theta \sim 0$ of the scattering amplitudes~(\ref{ScattAmp}) to a Schr\"odinger equation with a delta function potential. One then sees that a shallow bound state is formed for $\p<1$ in the $C$-even, adjoint channel. Its mass to leading order at large $\N$ is given by
\beq\label{MassAdj}
M_{\textrm{Adj}} = 2m\left(1-{\pi^2(1-\p)^2 \over 8\N^2\p^2} + \ldots\right)\, .
\eeq
We notice that it is at threshold when $\p=1$.  The approximation breaks down when $\p$ becomes too small. This is not completely surprising since the range of the potential increases as $\sim 1/\sqrt{p}$ at small $p$. Eventually, the interaction becomes confining at $\p=0$, where one expects a rather different scaling for the mass of the bound states \cite{Witten:1978bc}.

Finally let us add a few comments on the large rapidity behavior of the scattering amplitudes. In this limit $t \sim -s \gg 1$ and $\theta \sim \log{(s/m^2)} \sim 4\pi \kappa/\N \gg 1$. While $u_2 (\theta)$ and $r_1(\theta)$ both scale in this case as $\sim 1/\kappa$, this is not the case for $u_1(\theta)-1$, which has the following large rapidity asymptotics
\beq\label{AnBeh}
u_1(\theta) \simeq 1 -{i\pi \over \N \p} + \ldots \, .
\eeq
This is not exactly what one would expect for an asymptotically free theory and is apparently related to the fact that we have a circle with finite radius $r^2_\infty$ in the UV. Let us try to make this connection more precise. It is tempting to believe that the subleading large $\N$ corrections exponentiate in this regime
\beq\label{u1LR}
u_1(\theta) = e^{-i\pi/\N \p} + O(1/\theta)\,.
\eeq
For $\p=1$ this is in agreement with the large rapidity behavior of the finite $\N$ minimal reflectionless S-matrix, see Appendix~\ref{Smatrix}. The expression \eqref{u1LR} also coincides with the exact S-matrix \cite{Wiegmann:1985jt} for $\N=2$. Thus the equation \eqref{u1LR} seems to be the right guess. If we now re-express \eqref{u1LR} in terms of the Lagrangian parameter $r^2_{\infty} = 2\p\N$ we obtain $u_1(\theta) \simeq \exp{(-2i\pi/r_{\infty}^2)}$. This expression is reminiscent of the large rapidity behavior for the soliton scattering phase in the sine-Gordon model.%
\footnote{To make this rigorous one must take into account the proposal of \cite{Klassen:1992eq} that the sine-Gordon soliton scattering phase has to be normalized as $S(\theta=0) = -1$. This has the effect of multiplying the Zamolodchikov's S-matrix~\cite{Zamolodchikov:1978xm} by a minus sign.}
Notice that to match the normalization used in this paper one should consider sine-Gordon theory for a field $\varphi$ with period $\sqrt{\pi}$. The radius would then be given by $r^2_\infty = \beta^2/(2\pi)$ with $\beta^2$ being the sine-Gordon coupling. The ``anomalous'' behavior \eqref{AnBeh} then seems to indicate that the spinon in this theory is nothing else but a soliton for the $\varphi$ field. This is in line with the discussion in~\cite{Witten:1978bc} and this analogy will appear helpful when considering the theory in finite volume.

\section{Free energy computation}\label{MBA}

Having found hints of integrability at large $\N$, we now wish to verify whether integrability is present at finite values of $\N$. As we pointed out in the introduction, this is of particular interest from the viewpoint of the AdS/CFT correspondence, which requires considering $\N=4$. According to the analysis of~\cite{Koberle:1987wc, Campostrini:1993fr} and to our previous discussion, the theory at $\p=1$ is described by the minimal reflectionless $U(\N)$ S-matrix \cite{Berg:1977dp}. In this section we will check whether this assertion holds true at finite $\N$ and whether there are any modifications to the integrability condition $\p=1$. In order to do this we shall proceed with calculating the free energy of the theory at finite chemical potential. This can be done in two different ways. One can perform a perturbative computation to the desired order, or make use of the conjectured S-matrix. Compatibility of these two computations will allow us, with certain degree of confidence, to argue that integrability is present for finite $\N$. As a by-product of our analysis, we will be able to confirm that the value of $\p$, which will be kept arbitrary in the perturbative calculation, needs indeed to be set to $\p=1$ independently of the value of $\N$. 

The computation performed in this section is a standard analysis for two-dimensional integrable QFTs. It was introduced in the seminal papers~\cite{Hasenfratz:1990zz, Hasenfratz:1990ab} on the $O(\N)$ sigma models and numerous applications to a variety of integrable theories followed, see e.g. \cite{Forgacs:1991nk, Evans:1994sv}. The underlying idea is to consider the ground-state energy density $\varepsilon$ of a gas of spinons with a finite density $\rho$. The free energy of interest can then be obtained by a Legendre transformation
\beq
f(h) = \varepsilon(\rho) - \rho h\, ,
\eeq
with the chemical potential $h = d\varepsilon/d\rho$. Computing this quantity directly from the QFT leads to an expression in terms of the two RG-invariants of the theory, i.e., the dynamical scale $\Lambda$ and the deformation parameter $\p$,
\beq\label{fLambda}
f(h) = f(h, \Lambda, \p)\, .
\eeq
Thanks to asymptotic freedom, this analysis is tractable at large chemical potential, i.e., for $h\gg \Lambda$, where the theory is weakly coupled. On the other hand, the Bethe ansatz calculation, based on the asymptotic S-matrix, results in the free energy density directly in terms of the mass gap $m$ and arbitrary value of the chemical potential
\beq\label{fBA}
f(h) = f(h, m)\,.
\eeq
The Bethe ansatz computation will only lead to correct results at those values of $\p$ for which the S-matrix is a valid physical description of the model. This will be a single value for a generic value of $\N$, while in the case of $\N=2$ the analysis may be done for any $\p$, see \cite{Balog:1999ik}. The two computations \eqref{fLambda} and \eqref{fBA} should match at large chemical potential provided that
\beq\label{xidef}
m = \xi\Lambda\, ,
\eeq
where $\xi$ is a scheme-dependent constant. Since $\p$ enters the perturbative computations, it will also get fixed when both computations are compared. 

\subsection{Quantum field analysis}

For convenience we perform the QFT analysis using the bosonic formulation \eqref{BMdef} of the $U(1)\times SU(\N)$ model. Our starting point is the classical state describing the gas of spinons at large density. It is given by the single-spin classical solution
\beq\label{SSsol}
z_1 = e^{-i\omega \tau} \, , \qquad z_j = 0\, , \qquad j = 2, \ldots , \N\, ,
\eeq
where $\tau$ is the world-sheet time coordinate and $\omega$ a frequency that we shall soon relate to the charge density. This state is interesting because it is directly sensitive to the $U(1)$ sector of the theory and hence to the coupling $r^2 = 4\pi \kappa(1-\eta)$. It has the following energy and charge density
\beq
\varepsilon = \kappa(1-\eta)\omega^2\, , \qquad \rho = 2\kappa(1-\eta)\omega\, .
\eeq
We notice that in the $CP^{\N-1}$ limit $\eta \rightarrow 1$, both quantities vanish. This may be easily understood. In the $CP^{\N-1}$ model the state~(\ref{SSsol}) is equivalent to the Goldstone vacuum up to a gauge transformation and has therefore zero energy. 
For the free energy density we have
\beq
f(h)  = -\kappa(1-\eta)h^2 = -{r^2h^2 \over 4\pi}\, ,
\eeq
with the chemical potential $h=d\varepsilon/d\rho = \omega$.

Let us proceed now to the one-loop analysis. It is convenient to go to the Euclidean space $\tau \rightarrow -i\sigma_0$. To compute quantum corrections to the free energy, we shall expand the Lagrangian around the solution \eqref{SSsol} and calculate the partition function of the theory $\mathcal{Z}(h)$, which will allow us to extract the free energy
\beq
f(h) = -\frac{\log{\mathcal{Z}(h)}}{\textrm{volume of 2d space}}\,.
\eeq
A convenient parameterization around the background solution \eqref{SSsol} is given by
\beq\label{ParaSS}
z_1 = e^{-h  \sigma}{e^{i\vartheta} \over \sqrt{1+\bar{\chi}\chi}}\, , \qquad z_j = e^{-\alpha h  \sigma}{e^{i\vartheta}\chi_j \over \sqrt{1+\bar{\chi}\chi}} \, , \qquad j=2, \ldots , \N\, ,
\eeq
where $h  \sigma = h_{\mu}\sigma_{\mu}$, with $h_{\mu} = (h, 0)$ and $\sigma_{\mu} = (\sigma_{0}, \sigma_{1})$ the Euclidean world-sheet coordinates. This parametrization breaks the symmetry down to $U(\N-1)\times U(1)$. Note that we used that $\omega =h$ to have the chemical potential as a ``background field'' explicitely. More importantly, we have also allowed for a chemical potential $\alpha h$ for the $\N-1$ coordinates $z_{j\neq 1}$. This does not change the semiclassical limit since perturbatively $\chi_j \sim 1/\sqrt{\kappa} \sim 0$. It introduces however a new parameter $\alpha$, which eventually must be eliminated. This will be done by demanding that the system is in its ground state at fixed $h$. To see how it can be done, we plug the expressions \eqref{ParaSS} into the Lagrangian and expand up to quadratic order. We find up  to total derivatives
\beq
\Lag = -\kappa(1-\eta)h^2 + \Lag_2 + \ldots \, ,
\eeq
with
\beq
\Lag_2 = \kappa \left(\partial_\mu +\beta h_{\mu} \right)\bar{\chi} \left(\partial_\mu -\beta h_{\mu}\right) \chi + \kappa(1-\eta)^2h^2 \bar{\chi} \chi + \kappa(1-\eta)\partial_\mu \vartheta\partial_\mu \vartheta\, .
\eeq
The new parameter $\beta \equiv \alpha-\eta$. We see that the fields $\chi_j$ have acquired a mass $(1-\eta)h$, while the field $\vartheta$ has remained massless. The fields $\chi_j$ further couple to the chemical potential $\beta h_\mu$. For $\beta \neq 0$ the system therefore forms a condensate of massive bosons at rest. Minimizing the energy then amounts to choosing $\beta = 0$, or $\alpha = \eta$. The fields $\chi_j$ become free and their contribution to the free energy density is easily evaluated. The massless field does not contribute if one uses dimensional regularization.

The evaluation of the one-loop determinant for the $\N-1$ complex bosons is straightforward. It leads to
\beq\label{FEDr}
f(h) = -{r^2 h^2 \over 4\pi}\bigg[1+{2(\N-1)r^2 \over R^4}\left(\log{\left(r^2h \over R^2\mu\right)}- {1\over 2}\right) + O(1/R^6)\bigg]\, ,
\eeq
when expressed in terms of the radii. Along the way we absorbed the UV divergences into the renormalization of the radius $r$,
\beq
r^2 + {(\N-1)r^4 \over R^4}\left({2\over D-2} - \log{4\pi}+\gamma_{\textrm{E}}\right)\rightarrow r^2\, ,
\eeq
here performed in the $\overline{\textrm{MS}}$ scheme and with $\gamma_{\textrm{E}}$ the Euler-Mascheroni constant. This redefinition is in agreement with the renormalization group equations~(\ref{RGEBM}) derived from~\cite{Azaria:1995wg}. This is most easily seen at the level of the free energy density~(\ref{FEDr}). The latter defines a physical observable and its dependence on $\mu$ should drop out when the couplings $R^2, r^2$ fulfill the RG equations~(\ref{RGEBM}). This is easily verified to be the case, at the given order in perturbation theory. We can also check the consistency of the expression~(\ref{FEDr}) at special points. In the $O(2\N)$ limit, when $r^2 = R^2$, we obtain
\beq
f_{O(2\N)}(h) =  -{R^2 h^2\over 4\pi} \bigg[1+{2(\N-1) \over R^2}\left(\log{\left(h \over \mu\right)}- {1\over 2}\right) + O(1/R^4)\bigg]\, .
\eeq
This result agrees with the one of \cite{Hasenfratz:1990ab} when the couplings are appropriately related. In the opposite limit, $r^2 \rightarrow 0$, we find that $f(h)$ vanishes, as expected for the $CP^{\N-1}$ model.

To compare with the S-matrix computation one needs to express the free energy density \eqref{FEDr} in terms of the RG-invariant parameters of the model, i.e., in terms of the dynamical scale $\Lambda$ and deformation parameter  $\p$. Using the expressions \eqref{R2} and \eqref{r2} we arrive at
\beqa
\nn
f(h) &=& -{\p h^2\N \over 2\pi}\bigg[1-{\p(\N-1) \over \N\log{(h/\Lambda)}}-  \\
&&{\p(\N-1)(\N+\p-2)\log{\log{(h/\Lambda)}} + \p\N(\N-1)d_{\N} \over \N^2\log^2{(h/\Lambda)}} + \ldots\bigg]\, ,
\eeqa
now understood as an expansion valid at large chemical potential $h\gg \Lambda$. The constant $d_{\N}$ reads 
\beq\label{dN}
d_{\N} = -\log{\p} +{1-2\p\over 2}+{3\p-2\over 2\N}\, .
\eeq

\subsection{Matching the S-matrix analysis}\label{MMG}

To calculate the expression for the free energy of gas of spinons using the conjectured S-matrix, we first observe that only the spinon-spinon scattering phase in the symmetric channel, $S(\theta) = u_1(\theta)+u_2(\theta)$, is relevant for that purpose. The explicit expression for $S(\theta)$ may be found in Appendix~\ref{Smatrix}. It turns out that $S(\theta)$ is identical to the scattering phase for fundamental excitations in the $SU(\N)$ chiral Gross-Neveu model, up to the substitution $1/\N\rightarrow 1-1/\N$. The computation of the free energy density in the $SU(\N)$ chiral Gross-Neveu model was carried out in \cite{Forgacs:1991nk}. One can thus directly translate the result to the case at hand. For $h \gg m$ we find
\beq\label{SSBA}
f(h)  = -{h^2 \N \over 2\pi}\left[1- {\N-1 \over \N\log{(h/m)}} - {(\N-1)^2\log{\log{(h/m)}} + \N(\N-1)D_{\N} \over \N^2\log^2{(h/m)}} + \ldots\right] \, ,
\eeq
where $m$ is the mass of a spinon. The constant $D_{\N}$ stands for
\beq\label{DN}
D_{\N} = \log{\Gamma(1+1/\N)} -{\log{2} \over \N} -{1\over 2} + {3\over 2\N}\,.
\eeq

We immediately notice that the logarithmic pattern of the QFT and Bethe ansatz expressions, Eq.~(\ref{FEDr}) and~(\ref{SSBA}) respectively, are compatible. They match precisely if and only if $\p = 1$ for any $\N$. It confirms that integrability occurs at $\p=1$ without any $1/\N$ corrections. One other inference that may drawn is that the exact S-matrix has been properly identified, else the structures of both expansions would be different. Fixing further terms allows us to relate the mass gap $m$ to the dynamical scale $\Lambda$, i.e., determine the constant $\xi$ in~\eqref{xidef}. It is given by
\beq\label{xi}
\xi = \frac{m}{\Lambda}={(2/e)^{1/\N} \over \Gamma(1+1/\N)}\, .
\eeq
Note that this formula is valid in the $\overline{\textrm{MS}}$ scheme. At large $\N$ we immediately verify that $m = \Lambda +O(1/\N)$, as required by the large $\N$ analysis. For $\N=2$, the relation between $m$ and $\Lambda$ is known for any $\p$~\cite{Balog:1999ik}
\beq\label{mBalog}
m_{\N=2}(\p) = (2/e)^{1-\p/2} {4\Gamma(1+\p/2) \over \pi \p}\Lambda\, .
\eeq
It is easily seen to agree with \eqref{xi} when $\p=1$. We note also that for $p=2$ the mass gap~\eqref{mBalog} is identical to the one of the SUSY $CP^{1}$ model \cite{Evans:1994sv}, in line with the discussion in Section~\ref{sec:PCF}.

Finally, we observe that the expression~(\ref{xi}) is in agreement with the result reported in~\cite{Campostrini:1993fr}. This reference did not include any details on its derivation however. Moreover, it apparently overlooked the renormalization of the radius $r$, as pointed out in \cite{Azaria:1995wg}. The above analysis shows that the running of this coupling is actually necessary for a proper match of the QFT and S-matrix computations. We notice nevertheless that this feature is specific to our choice of probe. The computation of the constant $\xi$ could also be done by performing the one-loop analysis around a different classical solution, for instance the two-spin solution $\sim z_{1}^{J}\bar{z}_{\N}^{J}$. The free energy corresponding to the latter solution \textit{would not} be sensitive to the renormalization of the radius $r$, which could therefore be treated as a constant.  

\section{The AdS/CFT effective model}\label{BSM}

In this section we analyze in more details the particular case of the $CP^3$ model coupled to a Dirac fermion with charge $\e=2$, which was proposed in~\cite{Bykov:2010tv} to govern the low-energy effective theory of the Gubser-Klebanov-Polyakov (GKP) string~\cite{Gubser:2002tv} in the long-string limit.  For reader's convenience we recall the form of the Lagrangian
\beq\label{SM}
\Lag =\ka(\pa_{\mu}-iA_\mu)\bar{z} (\pa^{\mu}+iA^\mu) z + i\psib\gamma^{\mu}(\pa_{\mu}-2iA_\mu)\psi + \frac{1}{4\ka} \left(\psib \gamma_\mu \psi \right)^2\,.
\eeq
According to our previous analysis the fermonic model~\eqref{fermionicmodels} is integrable when the UV value of the Thirring coupling fulfills
\beq
\lambda_{\infty} =  \pi (\e^2/\N -1) = 0\, ,
\eeq
where in the last equality we have used the parameters of the string theory model. Please note that the absence of the UV value of $\lambda$ goes in line with the form of \eqref{SM}. The running of the Thirring coupling was found in \eqref{RGEbis} to be given by
\beq
\lambda = \lambda_{\infty} - {\e^2\over 2\N\kappa} + \ldots=-\frac{1}{2\kappa}+\ldots\, ,
\eeq
which upon plugging back into \eqref{fermionicmodels} leads immediately to \eqref{SM}! It is satisfying to find that the Lagrangian~(\ref{SM}), derived from a renormalizable and UV complete superstring action, displays the correct one-loop induced value of $\lambda$, which is scheme independent. We may thus conclude that \eqref{SM} is integrable and belongs to the class of integrable $U(1)\times SU(N)$ models. This corroborates the conjecture that the original $AdS_4 \times CP^3$ string theory sigma model is quantum integrable. Since this sigma model has a gauge theory dual and a set of all-loop spectral equations was conjectured for the said duality \cite{Gromov:2008qe}, we will independently propose Bethe equations for \eqref{SM}. 

\subsection{Physics in finite volume}

In this subsection we consider the fermionic model on a cylinder of length $L$. We want to address the problem of constructing the spectrum in the asymptotic domain $mL\gg 1$, where $m$ is the  mass of a spinon. To do that we will need to elucidate the role played by the fermion charge $\e$. We will assume it to be   integer, while the value of the Thirring coupling will be fixed requiring the integrability condition to be fulfilled. The spectral problem relevant to the string theory will then appear as a special case of $N=4$ and $\e=2$, and shall be discussed in more detail later on.

It is well known that for integrable theories the large volume spectrum is encoded in a set of asymptotic Bethe ansatz (ABA) equations. There are essentially two pieces of information required to write down such equations. The first ingredient is the S-matrix, which for the model at hand may be found in Appendix~\ref{Smatrix}. The second one is the choice of boundary conditions for the multi-spinon wave function. To minimize the amount of technical details, we will first look at scattering of $K$ spinons of the same $SU(\N)$ polarization. The S-matrix in this case reduces to the phase $S(\theta)$ introduced before. The corresponding ABA equations, which are the quantization conditions for the  momenta of the spinons, take the following form
\beq\label{ABAex}
e^{-ip(\theta_{k})L} = q\prod_{j\neq k}^{K}S(\theta_{k}-\theta_{j})\, , \qquad k=1, \ldots , K\, ,
\eeq
where $p(\theta) = m\sinh{\theta}$ is the momentum of a spinon with rapidity $\theta$. The phase $q$ is associated to the monodromy of the wave function as one of the spinons goes around the cylinder. Assuming that the bosonic fields $z^{i}$ of the theory are subject to periodic boundary conditions, one would naively choose $q=1$. We shall argue below that this is \textit{not} the proper choice for  generic values of $\e$.

So far the fermion charge $\e$ was subsumed into the parameter $\p$ and did not play a role on its own. We have, however, already mentioned its relation to a discrete $\mathbb{Z}_{\e}\cong \mathbb{Z}_{2\e}/\mathbb{Z}_2$ symmetry~\cite{Witten:1978bc}. Not surprisingly, understanding the effect of the charge $\e$ on the ABA equations parallels the implementation of this symmetry. To clarify this point we shall first develop a useful analogy with the $\e$-folded sine-Gordon (SG) model~\cite{Klassen:1992eq, Bajnok:2000wm}.

The latter model can be defined by the Lagrangian
\beq\label{efoldedSG}
\Lag_{\textrm{SG}} = {2\pi \e^2 \over \beta^2}\partial_\mu \varphi \partial^{\mu}\varphi + {m_0^2 \over \beta^2}\cos{(2\e\sqrt{\pi}\varphi)}\,,
\eeq
where $\varphi$ is a compact boson with period $\sqrt{\pi}$. Forgetting for a while the $SU(\N)$ symmetry of our problem, the model~(\ref{efoldedSG}) can be seen to have a lot in common with the model studied in this paper. They share for instance the same $\mathbb{Z}_{\e}$ symmetry which is generated by $\varphi \rightarrow \varphi + \sqrt{\pi}/\e$. Moreover, in both cases this symmetry is spontaneously broken in infinite volume. In the SG theory this is directly observable at the level of the Lagrangian \eqref{efoldedSG}. The theory has $\e$ degenerate vacua in a given period, which are located at $\varphi = n\sqrt{\pi}/\e$ with $n=0, \ldots , \e-1$. They are all equivalent and each breaks the discrete symmetry. In our case, the phenomenon is not visible at the tree level nor at any finite order in perturbation theory. It can be revealed however following the observation made in \cite{Witten:1978bc} that spinons are solitons from the perspective of the bosonized fermion $\varphi$. In the background of a spinon the field $\varphi$ jumps by
\beq
\Delta\varphi = {\sqrt{\pi} \over \e}Q\, ,
\eeq
where $Q=1$ is the spinon $U(1)$ charge. This immediately follows from the Gauss law~(\ref{BcFc}) upon bosonization, or equivalently from~(\ref{SolPhi}). Since the spinon is a stable massive particle carrying the minimal amount of $U(1)$ charge, one expects degenerate vacua separated from one another by $\Delta \varphi = \sqrt{\pi}/\e$. Up to inessential details like the explicit form of the potential or the expression for the soliton S-matrix, this picture is identical to the one emerging from~\eqref{efoldedSG}.

The charge $\e$ is associated to rescaling of the field $\varphi$ and may always be eliminated in infinite volume. In finite volume, on the other hand, the function of the charge $\e$ becomes somewhat more manifest. For definiteness we will assume periodic boundary conditions for the field $\varphi$, i.e., $\varphi(\sigma+L) = \varphi(\sigma)\, \, \textrm{mod}\, \, \sqrt{\pi}$. Due to tunneling effects the $\e$ vacua are no longer localized and their energy degeneracy is lifted leading to restoration of the $\mathbb{Z}_\e$ symmetry. Moreover, the entire Hilbert space of the theory splits into $\e$ subsectors associated to different representations of the $\mathbb{Z}_\e$ symmetry. Each subsector is associated to a Bloch wave with quasi-momentum $P_n = 2\pi n/\e$. The effect of a Bloch-wave background on the ABA equations is known~\cite{Zamolodchikov:1994uw, Bajnok:2000wm}: the wave function acquires an additional phase shift $q_n =\exp{(iP_n)}$ each time a soliton circles around the cylinder. An anti-soliton picks up the inverse phase $1/q_n$. One way of understanding it is by recalling \cite{Zamolodchikov:1994uw} that a Bloch-wave background is equivalent to inserting the vertex operator
\beq\label{Vn}
V_n \sim \exp{\left(i\e P_n \varphi/\sqrt{\pi} \right)}
\eeq
at the bottom of the cylinder. Under the shift $\varphi \rightarrow \varphi +\sqrt{\pi}/\e$ it transforms like
\beq
V_n \rightarrow q_n V_n\, ,
\eeq
as required for a Bloch wave with momentum $P_n$. This  background acts non-trivially on excitations with non-zero topological charge or winding number $W = \Delta\varphi/\sqrt{\pi}$ such that they pick up a phase $\exp{(i\e W P_n)}$ when transported once around $V_n$. This applies in particular to a soliton. The latter, interpolating between two adjacent vacua, contributes $W = 1/\e$  resulting in the phase shift $q_n$.

The overall effect of the charge $\e$ on the ABA equations is thus the division into subsectors characterized by a twist $q$, which itself is a root of unity
\beq\label{qeb}
q^{\e}=1\, .
\eeq 
There is a further modification~\cite{Bajnok:2000wm}. The  number of solitons minus antisolitons is quantized in the units of $\e$. This selection rule originates from the quantization of the total winding number $= (\varphi(\sigma)-\varphi(\sigma+L))/\sqrt{\pi}$. Since a soliton contributes only $1/\e$ unit to this number, at least $\e$ solitons are needed to make it integer.

The above remarks clarify why the charge $\e$ is related to the ABA equations. It is not yet clear however how exactly it modifies the spectrum of the fermionic theory. We shall see that equation \eqref{qeb} has to be slightly corrected. In fact equation \eqref{qeb} is pertinent to the $\mathbb{Z}_\e$ quotient of the bosonic model. This claim might seem puzzling given that we have argued before that the bosonic and fermionic models are equivalent. Nevertheless, examples of theories are known, which, despite being equivalent on the infinite plane, have different spectra in the finite volume. The most celebrated example is given by the duality between the sine-Gordon and the massive Thirring models~\cite{Klassen:1992eq}. The difference between the two sets of Bethe equations is tiny but significantly changes the spectra in finite volume. Here we are facing a similar problem.

There is a matter in which the bosonic and fermionic theories differ from each other. It is the action of the $\mathbb{Z}_2$ symmetry $\varphi \rightarrow \varphi + \sqrt{\pi}$, as mentioned in Section~\ref{FermMod}. This operation is trivial in the bosonic theory, where the field $\varphi$ is $\sqrt{\pi}$-periodic from the beginning. In the fermionic theory, however, this transformation is associated to the operator $(-1)^{F}$, with the fermion number
\beq
F = \textrm{number of fermions} - \textrm{anti-fermions}\,.
\eeq
The corresponding spaces of local (gauge-invariant) operators of the two theories are different due to the presence of operators anticommuting with $(-1)^F$ in the operator space of the fermionic theory. This is the root of the difference in the ABA equations for the two models. We refer the reader to~\cite{Klassen:1992eq} for a more detailed discussion in the context of the sine-Gordon / massive-Thirring duality.

We can now derive the correct twist $q$ for the spectral equations of the fermionic theory. We assume antiperiodic boundary conditions for the fermion $\psi$ while the bosons $z^{i}$ are taken to be periodic. There are two reasons for this particular choice. Firstly, we believe it is the proper set of boundary conditions to describe the string theory spectrum corresponding to $\N=4$ and $\e=2$. Secondly, the Neveu-Schwartz (NS) sector is the simplest one from the perspective of the state-operator mapping, which is a one-to-one correspondence between finite-volume eigenstates in the NS sector and local gauge-invariant vertex operators of the theory. These vertex operators are typically of the type
\beq
V \sim V_F V_B\, ,
\eeq
where $V_F$ and $V_B$ are written in terms of fermionic and bosonic fields, respectively. The bosonic part has a more transparent interpretation in terms of spinons. For example, the operator $V_B \sim z^{i_1} \ldots z^{i_K}$ is likely to correspond to a finite-volume state made out of $K$ spinons in a totally symmetric representation of $SU(\N)$. What needs to be understood is the effect of the fermionic component $V_{F}$ on these spinons. The fermionic operator has an electric charge which blots out the $U(1)$ charge of the bosonic operator $V_{B}$. This leads to a selection rule on the total number of spinons. We will come back later to this issue. The important point is that the vertex $V_F$ has also a ``Bloch-wave'' in close similarity to the operator $V_n$ considered in \eqref{Vn}. According to the rules of the bosonization, the operator $V_F$ with fermion number $F$ should contain the wave
\beq
\exp{\left(in\sqrt{\pi}\varphi \right)}\, ,
\eeq
where $n$ is an integer fulfilling $(-1)^n = (-1)^F$. By analogy with our previous discussion, we expect therefore that a spinon with winding number $1/\e$ will pick up an extra phase $\exp{(in\pi/\e)}$ due to the fermionic ``background''. This heuristic argument leads us to propose
\beq\label{qef}
q^{\e} = (-1)^{F}\, ,
\eeq
as the phase for the fermionic theory in the NS sector. We observe that for $F$ even it reduces to the twist~(\ref{qeb}) for the bosonic model. In other words, the bosonic and fermionic ABA  equations are the same for states which are neutral under the $\mathbb{Z}_2$ symmetry. This is analogous to what happens for the sine-Gordon / massive-Thirring duality \cite{Klassen:1992eq}, which in our notation corresponds to $\e=1$. We also notice that~(\ref{qef}) no longer defines a twist in the original sense since $q$ depends  via $F$ on the state considered. Finally, we should impose a selection rule for the fermionic theory. It may be traced back to the gauge invariance of the theory, which requires that the total bosonic $U(1)$ charge $Q=K-\bar{K}$, with $\bar{K}$ the total number of anti-spinons, is a multiple of the fermion charge $\e$. Explicitly,
\beq\label{SelRule}
K-\bar{K} = \e F\,.
\eeq
This selection rule is the same as for the bosonic theory, since in the NS sector $F \in \mathbb{Z}$.

\subsection{ABA equations}\label{sec:ABAeq}

We are now in the position to present the complete set of ABA equations for the NS sector of the string model~(\ref{SM}). To resolve the mixing related to the $SU(4)$ symmetry, one has to diagonalize the monodromy matrix associated to the minimal reflectionless S-matrix. The analysis can be done by means of the algebraic Bethe ansatz and is very similar to the one performed for the alternating Heisenberg spin chain of the planar ABJM theory \cite{Minahan:2008hf}. We refer the reader to this reference and here quote only the final result
\beq
\begin{aligned}\label{ABAeq}
e^{-ip(\theta_k) L} &= q\prod_{j\neq k}^{K}S(\theta_k-\theta_j)\prod_{j=1}^{\bar{K}}t_{1}(\theta_k-\bar{\theta}_j)\prod_{j=1}^{K_{1}}{2\theta_k/\pi -u_{1, j} +i/2 \over 2\theta_k/\pi -u_{1, j} -i/2}\, , \\
\prod_{j=1}^{K}{u_{1, k}-2\theta_{j}/\pi+\ft{i}{2} \over u_{1, k}-2\theta_{j}/\pi-\ft{i}{2}} &= \prod_{j\neq k}^{K_{1}}{u_{1, k}-u_{1, j}+i \over u_{1, k}-u_{1, j}-i} \prod_{j=1}^{K_{2}}{u_{1, k}-u_{2, j}-\ft{i}{2} \over u_{1, k}-u_{2, j}+\ft{i}{2}}\, , \\
1 &= \prod_{j\neq k}^{K_{2}}{u_{2, k}-u_{2, j}+i \over u_{2, k}-u_{2, j}-i} \prod_{j=1}^{K_{3}}{u_{2, k}-u_{3, j}-\ft{i}{2} \over u_{2, k}-u_{3, j}+\ft{i}{2}}\prod_{j=1}^{K_{1}}{u_{2, k}-u_{1, j}-\ft{i}{2} \over u_{2, k}-u_{1, j}+\ft{i}{2}}\, , \\
\prod_{j=1}^{\bar{K}}{u_{3, k}-2\bar{\theta}_{j}/\pi+\ft{i}{2} \over u_{3, k}-2\bar{\theta}_{j}/\pi-\ft{i}{2}} &= \prod_{j\neq k}^{K_{3}}{u_{3, k}-u_{3, j}+i \over u_{3, k}-u_{3, j}-i}\prod_{j=1}^{K_{2}}{u_{3, k}-u_{2, j}-\ft{i}{2} \over u_{3, k}-u_{2, j}+\ft{i}{2}}\, , \\
e^{-ip(\bar{\theta}_k) L} &= 1/q\prod_{j\neq k}^{\bar{K}}S(\bar{\theta}_{k}-\bar{\theta}_{j})\prod_{j=1}^{K}t_1(\bar{\theta}_{k}-\theta_{j})\prod_{j=1}^{K_{3}}{2\bar{\theta}_{k}/\pi - u_{3, j}+\ft{i}{2} \over 2\bar{\theta}_{k}/\pi - u_{3, j}-\ft{i}{2}}\, . \\
\end{aligned}
\eeq

In these equations $S(\theta)$ and $t_1(\theta)$ are the spinon-spinon and spinon-antispinon scattering phases, see Appendix \ref{Smatrix} for further details. The momentum of a spinon or anti-spinon, carrying respectively rapidity $\theta$ or $\bar{\theta}$, is given by $p(\theta) = m\sinh{\theta}$ and similarly for the anti-spinon. The total energy is given by
\beq
E = \sum_{k=1}^{K}m\cosh{\theta_k} + \sum_{k=1}^{\bar{K}}m\cosh{\bar{\theta}_k}\, . 
\eeq
The numbers $K_{1, 2, 3}$  count the isotopic roots $u_1, u_2$ and $u_3$, which change the flavors of the $K$ spinons and $\bar{K}$ anti-spinons, as may be inferred by looking at the $\mathfrak{su}(4)$ Dynkin labels of the state
\beq
[q_1, p, q_2] = [K-2K_{1} +K_{2}, K_{1}+K_{3}-2K_{2}, \bar{K}-2K_{3}+K_{2}]\, .
\eeq

The twist $q$ is solution of
\beq\label{q2Eq}
q^2 = (-1)^{F}\, ,
\eeq
which, together with \eqref{SelRule}, allows one to write
\beq
q = \pm \, \exp{(i\pi(K-\bar{K})/4)}\, .
\eeq
We stress that \textit{both} solutions have to be considered. In the $F$-odd sector this is required by the parity invariance of the theory, i.e., if $(\theta_k, \bar{\theta}_k, u_{i, k})$ is a solution of the ABA equations then so is $(-\theta_k, -\bar{\theta}_k, -u_{i, k})$. This property is not guaranteed if $q\neq 1/q$. Finally, we notice that an immediate consequence of the ABA equations is that the total momentum $P = \sum_{k}p(\theta_k)+\sum_k p(\bar{\theta_k})$ satisfies
\beq
e^{-iPL} = q^{K-\bar{K}} = (-1)^F\, .
\eeq
This is in agreement with our choice of boundary conditions for the string model.

\subsection{State-operator matching}

This subsection is devoted to investigating in more detail the mapping between solutions to the ABA equations~(\ref{ABAeq}) and local operators of the theory. According to this correspondence a state of energy $E$ and momentum $P$ is associated to a vertex operator of scaling dimension $\Delta$ and spin $S$. This relation should become more and more evident at small length $mL\ll 1$, for which~\cite{Cardy:1984rp, Affleck:1986bv, Bloete:1986qm}
\beq\label{Cardy}
E = -{\pi c \over 6L} + {2\pi \Delta \over L} + o(1/L)\, , \qquad P= {2\pi S \over L}\, ,
\eeq
with the UV central charge $c=2\N-1 = 7$. The second equality is actually expected to be valid for any $L$. The difficulty that one immediately encounters at the attempt of verifying \eqref{Cardy} with the help of \eqref{ABAeq} is that the ABA equations are only approximate. They do not take into account off-shell processes like vacuum tunneling, exchange of virtual particles, etc., which become more and more important at smaller values of $L$. As a consequence, the ABA equations often fail to reproduce~(\ref{Cardy}) making the mapping between operators and states hard to quantify. Despite these complications we will present some evidence that the twisted ABA equations correctly capture the properties of the low-lying energy eigenstates.

Let us start with the two ``vacua''. These states are easily found: they correspond to no root at all and have exactly zero energy in the ABA description. This is a trivial observation but it is nevertheless in agreement with the expectation that at large volume the vacua are exactly degenerate. The degeneracy should be lifted by tunneling processes as in the folded sine-Gordon theory~\cite{Zamolodchikov:1994uw, Bajnok:2000wm}. This is confirmed by the L\"uscher formula, which corrects the ABA equations at large volume $mL \gg 1$. With the twist $q$ included  the finite-size correction writes
\beq\label{Luescher}
E_{q} \simeq -{4q\over \pi}K_1(mL)-{4\over \pi q}K_1(mL)\, .
\eeq
Here, $K_1(z)$ is the modified Bessel's function with the asymptotics $K_1(z) \sim e^{-z}$ for  $z\gg 1$. We derived~(\ref{Luescher}) by adapting to our case the analysis performed  for the Bloch wave vacua of the sine-Gordon theory \cite{Zamolodchikov:1994uw} such that the $\N=4$ flavors of spinons and anti-spinons are correctly incorporated. The two NS vacua correspond to the two possible choices $q=\pm1$. The formula~(\ref{Luescher}) shows that the degeneracy is lifted and that the true vacuum has $q=1$. The NS ground state is thus $\mathbb{Z}_2$ even, as expected. It corresponds to the identity in the operator picture. But what is the vertex operator that creates the uplifted vacuum? It has to be odd under the $\mathbb{Z}_2$ symmetry $\varphi \rightarrow \varphi + \sqrt{\pi}/2$ and singlet under $SU(4)$. Moreover, among all such operators, it is likely to be the one with minimal scaling dimension $\Delta$ in order to minimize the energy gap at small volume
\beq\label{CF}
E_{-} - E_{+} = {2\pi \Delta \over L} + \ldots\, .
\eeq
There exist two possible candidates,
\beq\label{SecondVacuum}
\bar{\psi}\psi \sim \cos{(2\sqrt{\pi}\varphi)}\, , \qquad \textrm{and} \qquad \bar{\psi}\gamma_{5}\psi \sim \sin{(2\sqrt{\pi}\varphi)}\,.
\eeq
They are both spinless and have dimension $\Delta = 1$ to leading order at weak coupling.%
\footnote{Perturbative corrections to the scaling dimension are irrelevant here. Thanks to asymptotic freedom they are associated to subleading contributions at small length.}
To decide between the two we will look at the bosonized forms of the operators and use some basic quantum mechanics. We see that only the first operator describes a state with a wave function peaked around the (expected) minima $\varphi = 0, \sqrt{\pi}/2$ of the potential. On the contrary, the second operator depicts a wave function localized at the tips of the potential and is more likely to be an excited state in infinite volume. In any case, one of the two operators, or a linear combination perhaps, has to be an excited state and should be embedded as a non-trivial solution in the $q=-1$ subsector of the ABA equations. Assuming a singlet of $SU(4)$, the simplest solution in this subsector has vanishing roots $\theta = \bar{\theta}=u_1 = u_2 = u_3 =0$. Interestingly, its energy in the ABA approximation is exactly $2m$ for any length $L$ leading to a suggestive interpetation: it may be seen as the $C$-odd singlet bound state at rest which we know is exactly at threshold. This interpretation is in line with the comments in~\cite{Klassen:1992eq} on the vertex operators for breathers in the sine-Gordon theory.

A more direct test of the state-operator correspondence is available if one looks into the non-singlet $U(1)$ sector. Due to the selection rule the simplest configurations correspond to $(K,\bar{K})=(2,0)$ and its charge conjugate. Their fermion numbers are $F=\pm1$, respectively, and the allowed values of the twist are $q=\pm i$ in both cases. The solutions to the ABA equations are easily constructed at small length since in this regime the rapidities are large and we only need the asymptotic expression for the scattering phase $S(\theta)$, see Appendix~\ref{Smatrix}. The solutions with $F=1$ and $q=\pm i$, which minimize the energy, are given to leading order by
\beq\label{MomSpin}
p_1 \simeq \pm {\pi \over 4L} \qquad p_2 \simeq \pm {3\pi \over 4L}\,.
\eeq
The total energy and momentum take the following values
\beq\label{EP}
E \simeq {\pi \over L}\, , \qquad P  = \pm {\pi \over L}\, .
\eeq
They should be compared with~(\ref{Cardy}) after subtracting the energy of the vacuum ($\simeq -\pi c/(6L)$) and using the labels of the associated vertex operators. These should have fermion number $F=1$, belong to the symmetric representation $[2, 0, 0]$ of $SU(4)$, and have minimal scaling dimension. There are precisely two gauge-invariant operators with these properties,
\beq\label{psizz}
\psi_{\pm}z^i z^j\, ,
\eeq
where $\psi_\pm$ are the spin $S=\pm 1/2$ components of the Dirac field. Their scaling dimension is $\Delta = 1/2$ to leading order at weak coupling, in perfect agreement with~(\ref{EP}) and~(\ref{Cardy}). It is amusing to note that the energy and momentum of the states~(\ref{EP}) come directly from the dimension and the spin of the Fermi field in the operator picture~(\ref{psizz}) while they are distributed over the two spinons in the ABA description~(\ref{MomSpin}). Similar comparisons can be performed by considering more excited operators of the type $\psi_{\pm}D_{\pm} \psi_{\pm} z^i z^j z^k z^l$, etc., with $D_\pm$ being the light-cone covariant derivatives. 

This concludes our tests of the twisted ABA equations~(\ref{ABAeq}). Our discussion provides a somewhat appealing, though not very strong, evidence for the correctness of these equations. To put the analysis on a firmer ground, it would be interesting to write down and analyze the full-fledged thermodynamical Bethe ansatz (TBA) equations for the string model in the NS sector. As opposed to the ABA equations, the TBA equations should be valid for any length $L$. An analysis of this type was performed for the $\e$-folded sine-Gordon theory in~\cite{Bajnok:2000wm}.

\section{Conclusions}\label{Concl}

Integrability is a rare and unique property of quantum field theories. It is usually difficult to prove it rigorously since this would require a good handle on the non-perturbative physics of the theory. There are however certain quantities and features, one can study using perturbation theory or some non-perturbative methods, that provide hints as to whether the quantum integrability is present.

In this paper we have studied a family of $U(1) \times SU(N)$ theories, which may be formulated either exclusively in terms of bosonic degrees of freedom or by coupling a massless fermion with non-zero self-interaction. Both formulations are equivalent quantum mechanically and are equipped with a continuous parameter $\p$. It is related to the UV values of the coupling constants of the bosonic and fermionic models by 
\eqref{bosonicp} and \eqref{pferm}, respectively. Since by definition it is RG-invariant, it can be deemed \textit{physical} at the quantum level. The corresponding $(\p,\N)$ family of models covers a wide range of theories, which are summarized in Table~\ref{table:CP}.

We have found several indications that the class of models considered may be rendered integrable with the right choice of the physical parameter. This fine-tuning is not necessary for $\N=2$ because the $U(1) \times SU(2)$ model is equivalent to an integrable one-parameter deformation of the PCF model. Our analysis suggests that the theory should remain integrable at higher values of $\N$ at least for $\p=1$. This may be inferred already at the level of the large $\N$ scattering matrix, where integrability places severe constraints on its structure. A similar conclusion should be reached by studying the $2 \to 4$ amplitude at leading order in large $\N$. An important question is whether this choice of $\p$ receives any modification beyond the large $\N$ limit. The one-loop free energy computation in Section~\ref{MBA} is a direct confirmation that this is not the case. This leads us to believe that the $U(1)\times SU(\N)$ model is integrable for $\p=1$. We should stress however that this might not be the only point where integrability prevails. Alternative finite values of $\p$ might exist, though presumably only at small $\N$, and it would be of interest to investigate more closely this eventuality.

A model belonging to the integrable class identified in this paper has been recently found in \cite{Bykov:2010tv}. It governs the dynamics of massless excitations around the GKP solution of the $AdS_4 \times CP^3$ string theory. These excitations comprise the $CP^3$ degrees of freedom and a Dirac fermion. The Lagrangian of the model is embedded in the fermionic formulation \eqref{fermionicmodels} and its parameters fulfill the integrability condition $\p=1$. The integrability of this effective model is a strong evidence in favor of the integrability of the full $AdS_4 \times CP^3$ super sigma model. In Section \ref{BSM} we put forward a set of asymptotic Bethe equations for the fermionic model. Our argumentation relied solely on the physics of the model and was not inspired by the AdS/CFT correspondence. This set of spectral equations may thus be  used to test the veracity of the all-loop ABA equations proposed for the $AdS_4 / CFT_3$ duality \cite{Gromov:2008qe}.

\begin{table}[t]
\centering
\begin{tabular}{|l|p{3cm}|p{4.5 cm}|p{3cm}|}
  \hline
  & $\p=0$ & $0 < \p < \infty $ &$\p=\infty$ \\
  \hline
  \hline
  $\N=2$ & $O(3)$ model\, \, \, \, \, \, \, \,  +~free boson & $U(1)\times SU(2)$ model; \, \, \, \, \, \, \, \, integrability at any $\p$ & $O(4)$ model \\ 
  \hline 
  $\N>2$ & $CP^{\N-1}$ model\, \, \, \, \, \, \, \,  +~free boson & $U(1)\times SU(\N)$ model;\, \, \, \, \, \, \, \,  integrability at $\p=1$ & $O(2\N)$ model \\
  \hline
\end{tabular}
\caption{\label{table:CP} Different limiting cases of the model under consideration.}
\end{table}

\section*{Acknowledgments}
We would like to thank Juan Maldacena for discussions and valuable suggestions. One of us (B.B.) is most grateful to David Andriot, Andrei Belitsky, Dmitri Bykov, and Gregory Korchemsky, for helpful discussions. The research of Adam Rej was supported by a Marie Curie International Outgoing Fellowship within the 7th European Community Framework Programme, grant number PIOF-GA-2010-273854. Adam Rej also gratefully acknowledges support from the Institute for Advanced Study.

\appendix

\section{Renormalization of the Thirring coupling}\label{RTC}

In this appendix we show that the Thirring coupling runs. In order to do this, we will construct the two-point function for the fermionic current $j^{\mu} = \bar{\psi}\gamma_\mu \psi$ up to order $O(1/\kappa^2)$ in the weak coupling expansion. More precisely, we shall consider the form factor $D(p^2)$ defined by
\beq\label{Corr}
\int d^{D}x \left< j^{\mu}(x)j^{\nu}(0)\right> e^{ipx} = -{i \over \pi D(p^2)} \left(\eta^{\mu \nu} -{p^{\mu} p^{\nu} \over p^2}\right)\, .
\eeq
Notice that this quantity is dimensionless in two dimensions ($D=2$). An important property of this correlation function is that it is observable and hence UV finite. We will demonstrate that the finiteness cannot be preserved if the Thirring coupling does not renormalize.

We should first comment on the renormalization of the bosonic coupling $1/\kappa$ itself. At leading order the running of this coupling constant follows that of the $CP^{\N-1}$ model. The reason is that the contribution from a fermionic loop are controlled by the correlator (\ref{Corr}) evaluated in the Thirring model. The latter is well-known to be finite,
\beq
D(p^2)_{\textrm{Thirring}} = 1+\lambda/\pi + O(D-2)\, ,
\eeq
for $D=2$. Hence all divergent corrections to the coupling $1/\kappa$ come from bosonic loops at leading order. In the $\overline{\textrm{MS}}$ scheme the renormalized coupling $1/\kappa_{\textrm{ren}}$ is then defined as in the $CP^{\N-1}$ model
\beq\label{KappaRen}
\kappa_{\textrm{ren}} = \kappa + {\N \over 4\pi}\left[{2\over D-2} +\gamma_{\textrm{E}} - \log{4\pi}\right] + O(1/\kappa)\,.
\eeq
We refer the reader to~\cite{Valent:1984rj, Hikami:1979ih} for an explicit derivation of this relation and for a more detailed discussion of the renormalization of the $CP^{\N-1}$ model.

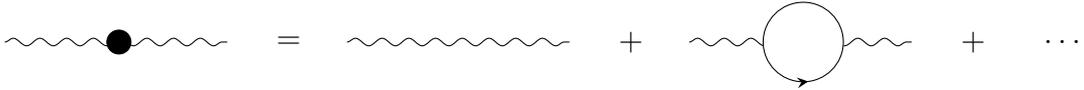
\begin{figure}[t]
\begin{center}
\begin{tikzpicture}[scale=1.5]
\draw[decorate, decoration={snake,amplitude=\amp}] (0,0) -- (1.95,0);
\draw[decorate, decoration={snake,amplitude=\amp}] (3,0) -- (4.95,0);
\draw[decorate, decoration={snake,amplitude=\amp}] (6,0) -- (7.95,0);
\draw[draw=white, fill=white] (7,0) circle (10pt);
\draw[draw=black,  decoration={markings, mark=at position 0.77 with {\arrow[thick]{stealth}}}, postaction={decorate}] (7,0) circle (10pt);
\draw[fill=black] (1,0) circle (3pt);
\node at (2.49,0) {$=$};
\node at (2.49+3,0) {$+$};
\node at (2.49+6,0) {$+$};
\node at (2.49+6.78,0) {$\ldots$};
\end{tikzpicture}
\end{center}
\caption{\label{Photon} Gauge field propagator to lowest order.}
\end{figure}

In order to calculate $D(p^2)$, we shall first derive the effective propagator for the gauge field $A_\mu$,
\beq
\int d^{D}x \, e^{ipx} \left< A^{\mu}(x)A^{\nu}(0)\right>_{CP^{\N-1}} = D^{\mu \nu}(p^2)\, ,
\eeq
by integrating out the bosonic degrees of freedom of the $CP^{\N-1}$ model. To leading order at weak coupling, this model is described by free massless complex bosons coupled to a gauge field. The relevant part of the Lagrangian is
\beq
\Lag = \kappa\partial_\mu \bar{\chi}\partial^{\mu}\chi - i\kappa A^{\mu}\bar{\chi}\overleftrightarrow{\partial_{\mu}}\chi + \kappa A_\mu A^{\mu} + \ldots\, ,
\eeq
where $\chi = (\chi_2, \ldots, \chi_{\N})$ is a multiplet of $\N-1$ complex bosons and $\bar{\chi}\overleftrightarrow{\partial_{\mu}}\chi = \bar{\chi}\partial_{\mu}\chi - \partial_{\mu}\bar{\chi}\chi$. Note that these fields are not subject to any constraint since they parameterize the transverse directions around the Goldstone vacuum $z = (1, 0, \ldots , 0)$. The contribution of these massless bosons to the propagator of the gauge field is depicted in Fig.~\ref{Photon}.
A direct computation of the one-loop diagram leads to
\beq\label{PhProp}
D^{\mu \nu}(p^2) = {i\eta^{\mu\nu} \over 2\kappa} - {\N-1\over 4\kappa^2}\mu^{2-D}\int{d^D q \over (2\pi)^{D}}{(p+2q)^{\mu}(p+2q)^{\nu} \over (p+q)^2q^2} + O(1/\kappa^3)\, ,
\eeq
where the scale $\mu$ was introduced to keep the coupling $\kappa$ dimensionless. A straightforward algebra yields
\beq
\mu^{2-D}\int{d^D q \over (2\pi)^{D}}{(p+2q)^{\mu}(p+2q)^{\nu} \over (p+q)^2q^2} = -{I(p^2) \over D-1}\left(\eta^{\mu \nu} -{p^{\mu} p^{\nu} \over p^2}\right)\, ,
\eeq
where
\beq\label{Ip}
I(p^2) =\int{d^D q \over (2\pi)^{D}}{ \mu^{2-D} p^2 \over (p+q)^2q^2} = {i(2\sqrt{\pi})^{3-D} \over 2^{D}\sin{(\pi D/2)}\Gamma(D/2-1/2)}\left({-p^2 \over \mu^2}\right)^{D/2-1}\, .
\eeq
Notice that this integral is divergent for $D=2$.

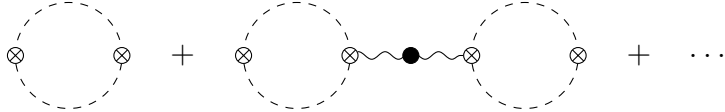
\begin{figure}
\begin{center}
\begin{tikzpicture}[cross/.style={path picture={ 
  \draw[black]
(path picture bounding box.south east) -- (path picture bounding box.north west) (path picture bounding box.south west) -- (path picture bounding box.north east);
}}]
\draw[dashed, postaction={decorate}] (0,0) circle (20pt);
\draw[dashed, postaction={decorate}] (3,0) circle (20pt);
\draw[dashed, postaction={decorate}] (6,0) circle (20pt);
%\draw[dashed,decoration={markings, mark=at position 0.77 with {\arrow[thick]{stealth}}}, postaction={decorate}] (0,0) circle (20pt);
%\draw[dashed, decoration={markings, mark=at position 0.77 with {\arrow[thick]{stealth}}}, postaction={decorate}] (3,0) circle (20pt);
%\draw[dashed, decoration={markings, mark=at position 0.77 with {\arrow[thick]{stealth}}}, postaction={decorate}] (6,0) circle (20pt);
\draw[{decorate, decoration={snake,amplitude=\amp}}] (0.7+3,0)--(-0.7+6,0);
\draw[fill=black] (4.5,0) circle (3pt);
\draw[draw=white, fill=white] (-0.7,0) circle (3pt);
\draw[cross] (-0.7,0) circle (3pt);
\draw[draw=white, fill=white] (0.7,0) circle (3pt);
\draw[cross] (0.7,0) circle (3pt);
\draw[draw=white, fill=white] (-0.7+3,0) circle (3pt);
\draw[cross] (-0.7+3,0) circle (3pt);
\draw[draw=white, fill=white] (0.7+3,0) circle (3pt);
\draw[cross] (0.7+3,0) circle (3pt);
\draw[draw=white, fill=white] (-0.7+6,0) circle (3pt);
\draw[cross] (-0.7+6,0) circle (3pt);
\draw[draw=white, fill=white] (0.7+6,0) circle (3pt);
\draw[cross] (0.7+6,0) circle (3pt);
\node at (1.5,0) {$+$};
\node at (7.5,0) {$+$};
\node at (7.5+0.9,0) {$\ldots$};
\end{tikzpicture}
\end{center}
\caption{\label{CurrentCurrent} Geometric series for the diagrams contributing to the divergent part of the current-current correlator. Crosses stand for insertions of the fermionic current. Dashed lines indicate contractions performed in the massless Thirring theory.}
\end{figure}

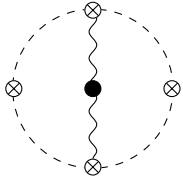
\begin{figure}
\begin{center}
\begin{tikzpicture}[cross/.style={path picture={ 
  \draw[black]
(path picture bounding box.south east) -- (path picture bounding box.north west) (path picture bounding box.south west) -- (path picture bounding box.north east);
}}]
\draw[dashed, postaction={decorate}] (0,0) circle (30pt);
\draw[{decorate, decoration={snake,amplitude=\amp}}] (0,1.04)--(0,-1.04);
\draw[draw=white, fill=white] (-1.04,0) circle (3pt);
\draw[cross] (-1.04,0) circle (3pt);
\draw[draw=white, fill=white] (0,1.04) circle (3pt);
\draw[cross] (0,1.04) circle (3pt);
\draw[draw=white, fill=white] (0,-1.04) circle (3pt);
\draw[cross] (0,-1.04) circle (3pt);
\draw[draw=white, fill=white] (1.04,0) circle (3pt);
\draw[cross] (1.04,0) circle (3pt);
\draw[draw=black, fill=black] (0,0) circle (3pt);
\end{tikzpicture}
\end{center}
\caption{\label{Irr} Example of a diagram contributing to the finite part of the correlator. It represents the connected part of the four-point function of fermionic currents in the massless Thirring theory convoluted with the gauge field propagator. Since the first one is of order $O(D-2)$, it compensates the pole in the latter $\sim 1/(D-2)$ leading to a finite contribution.}
\end{figure}

We are now in the position to compute the correction to the two-point function~(\ref{Corr}), or more precisely the divergent part thereof. We calculate it up to $O(1/\kappa^2)$ in the weak coupling expansion, but to all orders in the Thirring coupling. The relevant diagrams are shown in Figure \ref{CurrentCurrent}. Their contribution is a geometric series that sums up to
\beq\label{Dp}
D(p^2) = 1+ {\lambda \over \pi} + {\e^2 \over 2\pi \kappa} - {i\e^2(\N-1)I(p^2) \over 4\pi \kappa^2(D-1)}  + O(1/\kappa^3)\, .
\eeq
We stress that this computation may not be accurate enough to capture the finite part of the perturbative correction $\sim 1/\kappa^2$. The reason is that we are disregarding certain diagrams which are superficially of order $O(D-2)$ but might nonetheless make a finite contribution when convoluted with the divergent propagator~(\ref{PhProp}). An example of such diagrams is depicted in Figure \ref{Irr}.
Our analysis should reproduce correctly the pole at $D=2$ however. Taking the limit $D \to 2$ in \eqref{Dp} with help of \eqref{Ip} we get
\beq
D(p^2) = 1+ {\lambda \over \pi} + {\e^2 \over 2\pi \kappa} - {\e^2(\N-1) \over 4\pi^2(D-2)\kappa^2}  + \ldots\, .
\eeq
The result is UV divergent. The part linear in $\N$ will be made finite after renormalization of the bosonic coupling, see \eqref{KappaRen}. To absorb the remaining divergence we must renormalize the Thirring coupling,
\beq
\lambda + {\e^2 \over 8\pi \kappa^2}\bigg[{2\over (D-2)}+\gamma_{\textrm{E}} -\log{4\pi}\bigg] \rightarrow \lambda\, ,
\eeq
where we again work in the $\overline{\textrm{MS}}$ scheme. The resulting RG equation takes the form
\beq
\mu{\partial \lambda \over \partial \mu} = {\e^2 \over 4\pi \kappa^2} + O(1/\kappa^3)\, .
\eeq
This is the expression we used in Section \ref{FermMod}. It was found to be in agreement with the RG equation of the dual bosonic theory. The calculation performed here is not very different from the free energy density computation done for the bosonic model in Section~\ref{MBA}. Both lead to the same conclusion that the deformation parameter of the bosonic and fermionic model must be renormalized.

\section{Exact S-matrix}\label{Smatrix}

In this appendix we discuss some properties of the factorized $S$-matrices with $U(\N)$ symmetry. Their complete classification has been proposed long time ago by Berg et al.~\cite{Berg:1977dp}. It was found that there are several  classes of solutions and the minimal representatives of each class do not depend on continuous parameters for $\N > 1$. This immediately implies that there is no continuous deformation of the minimal $O(2\N)$ S-matrix~\cite{Zamolodchikov:1978xm} preserving the $U(1)\times SU(\N)$ symmetry.%
\footnote{The $O(2\N)$ S-matrix was found to be embedded in the class III of the classification, see discussion in~\cite{Berg:1977dp}.}
This important result can be used to argue that the integrability of the $U(1)\times SU(\N)$ models considered in this paper can only be sporadic with respect to the deformation parameter $\p$.

The absence of a continuous deformation of the $O(2\N)$ S-matrix is however in disagreement with the S-matrix derived by Wiegmann \cite{Wiegmann:1985jt} for $\N=2$. He found a solution given by a tensor product of the minimal $SU(2)$ S-matrix and the sine-Gordon S-matrix. The latter contains a free continuous parameter $p$. In this appendix we would like to clarify this apparent paradox. Contrarily to what has been stated in \cite{Berg:1977dp}, we show that the case $\N=2$ is special and permits the solution found by Wiegmann. The space of solutions for $\N=2$ is enlarged in the very same way as in the case of $O(\N)$ factorized S-matrices \cite{Zamolodchikov:1978xm}. Explicitly, we observe a reduction of the number of equations constraining the S-matrix due to additional identities inherent to $\N=2$. We have no doubt that this caveat has been known before. We did not find however any discussion nor comment in the literature, which is why we decided to comment on it below.

\subsection{The Yang-Baxter equations}

The equations that implement the factorization of the S-matrix are the so-called Yang-Baxter equations. Their construction is facilitated by the use of the Faddeev-Zamolodchikov algebra, see for instance \cite{Zamolodchikov:1978xm}. The spinon (antispinon) carrying rapidity $\theta$ is denoted by $Z_i(\theta)$ (respectively $\bar{Z}_i(\theta)$) with the $SU(\N)$ index $i=1, \ldots, \N$. The incoming states are written as strings of such symbols ordered with decreasing rapidities. An opposite ordering is used for the outgoing states. Their algebraic relations are encoded into the commutation relations between different $Z$'s and/or $\bar{Z}$'s. For example the spinon-spinon scattering is implemented by
\beq
Z_{i}(\theta_1)Z_{j}(\theta_2) = S_{ij}^{kl}(\theta)Z_{l}(\theta_2)Z_{k}(\theta_1) \, ,
\eeq
where $\theta=\theta_1 - \theta_2, \theta_1 \geqslant \theta_2,$ and $S_{ij}^{kl}(\theta)$ is the two-to-two spinon S-matrix. Similarly, for the spinon-antispinon scattering one has
\beq
Z_{i}(\theta_1)\bar{Z}_{j}(\theta_2) = F_{ij}^{kl}(\theta)\bar{Z}_{l}(\theta_2)Z_{k}(\theta_1)  +  B_{ij}^{lk}(\theta)Z_{k}(\theta_2)\bar{Z}_{l}(\theta_1)\,.
\eeq
The matrices $F$ and $B$ stand for the forward (transmitted) and backward (reflected) components of the scattering matrix, respectively. The corresponding formulas for $\bar{Z}$ are obtained by charge conjugation. 

Imposing the $U(\N)$ symmetry allows to decompose the matrices $S, F, B$ into six scalar amplitudes, as done in (\ref{SUnDec}). The crossing symmetry yields three linear relations between them, see (\ref{Crossing}). Unitarity, on the other hand, provides quadratic constraints. These relations are most easily written in terms of the amplitudes associated to the various $SU(\N)$ invariant channels~\cite{Berg:1977dp}. By way of illustration, after introducing the scalar factor for the scattering of two spinons in the symmetric channel,
\beq\label{SphaseSym}
S(\theta) =S_{11}^{11}(\theta) = u_1(\theta)+u_2(\theta)\, ,
\eeq
one arrives at the unitarity equation $S(\theta)S(-\theta) =1$. There are six such equations for the six $SU(\N)$-invariant channels of spinon-spinon and spinon-antispinon scattering. Their explicit expression can be found in \cite{Berg:1977dp}.

Finally, we have the Yang-Baxter equations. These are cubic in the $S, F, B$ matrices and take care of consistency of the factorization of the three-body scattering \cite{Zamolodchikov:1978xm}. They are equivalent to the associativity of the Faddeev-Zamolodchikov algebra. In our case we need to consider the four distinct processes with initial and final states listed below
\beq\label{AssocFZ}
\begin{aligned}
&Z_{i}(\theta_1)Z_{j}(\theta_2)Z_{k}(\theta_3) \rightarrow Z_{n}(\theta_3)Z_{m}(\theta_2)Z_{l}(\theta_1)\, , \\
&Z_{i}(\theta_1)Z_{j}(\theta_2)\bar{Z}_{k}(\theta_3) \rightarrow \bar{Z}_{n}(\theta_3)Z_{m}(\theta_2)Z_{l}(\theta_1)\, , \\
&Z_{i}(\theta_1)Z_{j}(\theta_2)\bar{Z}_{k}(\theta_3) \rightarrow Z_{n}(\theta_3)\bar{Z}_{m}(\theta_2)Z_{l}(\theta_1)\, , \\
&Z_{i}(\theta_1)Z_{j}(\theta_2)\bar{Z}_{k}(\theta_3) \rightarrow Z_{n}(\theta_3)Z_{m}(\theta_2)\bar{Z}_{l}(\theta_1)\, .
\end{aligned}
\eeq
The inequivalent ways of computing the S-matrix for these processes, obtained by iterative use of the commutation relations, yield cubic matrix identities between $S, F, B$. The equations corresponding to the first two lines in \eqref{AssocFZ} constrain the spinon-spinon scattering matrix $S$ and relate $F$ to $S$. They are oblivious to the backward scattering matrix $B$. Each of them provides a single functional relation,
\beq
f(\theta_{12}) + f(\theta_{23}) = f(\theta_{13})\, , \qquad h(\theta_{13}) - h(\theta_{23}) = -f(\theta_{12})\, ,
\eeq
where $\theta_{ij} = \theta_{i}-\theta_{j}, f(\theta) = u_1(\theta)/u_2(\theta),$ and $h(\theta) = t_1(\theta)/t_2(\theta)$. The general solution to these relations is well-known \cite{Zamolodchikov:1978xm, Berg:1977dp}. After taking into account the crossing relations between $u_{1,2}(\theta)$ and $t_{1,2}(\theta)$ it reads
\beq\label{ut}
u_1(\theta) = {i\theta \over \nu}u_2(\theta)  \, , \qquad t_1(\theta) = -{i \over \nu}\left(\theta-i\pi\right)t_2(\theta)\, ,
\eeq
where the parameter $\nu$ is still to be determined. The relations \eqref{ut} apply to all $U(1)\times SU(\N)$ factorized S-matrices which are of interest to us, that is assuming that $u_1(\theta), t_1(\theta) \neq 0$.

\begin{figure}
\hspace*{-6mm}
\begin{tabular}{x{5cm} x{0.1cm} x{5cm} x{0.1cm} x{5cm}}
\begin{tikzpicture}[thick, scale=2]
\draw[decoration={markings, mark=at position 0.55 with {\arrow{stealth}},mark=at position 0.08 with {\arrow{stealth}},mark=at position 0.96 with {\arrow{stealth reversed}}},
        postaction={decorate}] (0,0) -- (0,2);
\draw[decoration={markings, mark=at position 0.55 with {\arrow{stealth}},mark=at position 0.08 with {\arrow{stealth}},mark=at position 0.96 with {\arrow{stealth}}},
        postaction={decorate}] (-0.36,0.186) -- (1.36,1.21);
\draw[decoration={markings, mark=at position 0.55 with {\arrow{stealth}},mark=at position 0.08 with {\arrow{stealth reversed}},mark=at position 0.96 with {\arrow{stealth}}},
        postaction={decorate}] (-0.36,1.82) --(1.36,0.78);
\draw[fill=black!10] (0,1.6) circle (5pt);
\draw[fill=black!10] (1,1) circle (5pt);
\draw[fill=black!10] (0,0.4) circle (5pt);
\node[font=\fontsize{9}{9}] at (0,1.6) {$\textrm{B}$};
\node[font=\fontsize{9}{9}] at (1,1) {$\textrm{F}$};
\node[font=\fontsize{9}{9}] at (0,0.4) {$\textrm{S}$};
\end{tikzpicture} 
&
=
\vspace{1.8mm}
&
\begin{tikzpicture}[thick, scale=2]
\draw[decoration={markings, mark=at position 0.55 with {\arrow{stealth}},mark=at position 0.08 with {\arrow{stealth}},mark=at position 0.96 with {\arrow{stealth reversed}}},
        postaction={decorate}] (1,0) -- (1,2);
\draw[decoration={markings, mark=at position 0.55 with {\arrow{stealth reversed}},mark=at position 0.08 with {\arrow{stealth reversed}},mark=at position 0.96 with {\arrow{stealth}}},
        postaction={decorate}] (1.36,0.186) -- (-0.36,1.21);
\draw[decoration={markings, mark=at position 0.55 with {\arrow{stealth}},mark=at position 0.08 with {\arrow{stealth reversed}},mark=at position 0.96 with {\arrow{stealth reversed}}},
        postaction={decorate}] (1.36,1.82) --(-0.36,0.78);
\draw[fill=black!10] (1,1.6) circle (5pt);
\draw[fill=black!10] (0,1) circle (5pt);
\draw[fill=black!10] (1,0.4) circle (5pt);
\node[font=\fontsize{9}{9}] at (1,1.6) {$\textrm{B}$};
\node[font=\fontsize{9}{9}] at (0,1) {$\textrm{B}$};
\node[font=\fontsize{9}{9}] at (1,0.4) {$\textrm{F}$};
\end{tikzpicture} 
&
+
\vspace{1.8mm}
&
\begin{tikzpicture}[thick, scale=2]
\draw[decoration={markings, mark=at position 0.55 with {\arrow{stealth reversed}},mark=at position 0.08 with {\arrow{stealth}},mark=at position 0.96 with {\arrow{stealth reversed}}},
        postaction={decorate}] (1,0) -- (1,2);
\draw[decoration={markings, mark=at position 0.55 with {\arrow{stealth}},mark=at position 0.08 with {\arrow{stealth reversed}},mark=at position 0.96 with {\arrow{stealth}}},
        postaction={decorate}] (1.36,0.186) -- (-0.36,1.21);
\draw[decoration={markings, mark=at position 0.55 with {\arrow{stealth reversed}},mark=at position 0.08 with {\arrow{stealth reversed}},mark=at position 0.96 with {\arrow{stealth reversed}}},
        postaction={decorate}] (1.36,1.82) --(-0.36,0.78);
\draw[fill=black!10] (1,1.6) circle (5pt);
\draw[fill=black!10] (0,1) circle (5pt);
\draw[fill=black!10] (1,0.4) circle (5pt);
\node[font=\fontsize{9}{9}] at (1,1.6) {$\textrm{F}$};
\node[font=\fontsize{9}{9}] at (0,1) {$\textrm{S}$};
\node[font=\fontsize{9}{9}] at (1,0.4) {$\textrm{B}$};
\end{tikzpicture} 
\tabularnewline
\begin{tikzpicture}[thick, scale=2]
\draw[decoration={markings, mark=at position 0.55 with {\arrow{stealth}},mark=at position 0.08 with {\arrow{stealth}},mark=at position 0.96 with {\arrow{stealth}}},
        postaction={decorate}] (0,0) -- (0,2);
\draw[decoration={markings, mark=at position 0.55 with {\arrow{stealth}},mark=at position 0.08 with {\arrow{stealth}},mark=at position 0.96 with {\arrow{stealth reversed}}},
        postaction={decorate}] (-0.36,0.186) -- (1.36,1.21);
\draw[decoration={markings, mark=at position 0.55 with {\arrow{stealth reversed}},mark=at position 0.08 with {\arrow{stealth reversed}},mark=at position 0.96 with {\arrow{stealth}}},
        postaction={decorate}] (-0.36,1.82) --(1.36,0.78);
\draw[fill=black!10] (0,1.6) circle (5pt);
\draw[fill=black!10] (1,1) circle (5pt);
\draw[fill=black!10] (0,0.4) circle (5pt);
\node[font=\fontsize{9}{9}] at (0,1.6) {$\textrm{S}$};
\node[font=\fontsize{9}{9}] at (1,1) {$\textrm{B}$};
\node[font=\fontsize{9}{9}] at (0,0.4) {$\textrm{S}$};
\end{tikzpicture} 
&
=
\vspace{1.8mm}
&
\begin{tikzpicture}[thick, scale=2]
\draw[decoration={markings, mark=at position 0.55 with {\arrow{stealth}},mark=at position 0.08 with {\arrow{stealth}},mark=at position 0.96 with {\arrow{stealth}}},
        postaction={decorate}] (1,0) -- (1,2);
\draw[decoration={markings, mark=at position 0.55 with {\arrow{stealth reversed}},mark=at position 0.08 with {\arrow{stealth reversed}},mark=at position 0.96 with {\arrow{stealth}}},
        postaction={decorate}] (1.36,0.186) -- (-0.36,1.21);
\draw[decoration={markings, mark=at position 0.55 with {\arrow{stealth}},mark=at position 0.08 with {\arrow{stealth}},mark=at position 0.96 with {\arrow{stealth reversed}}},
        postaction={decorate}] (1.36,1.82) --(-0.36,0.78);
\draw[fill=black!10] (1,1.6) circle (5pt);
\draw[fill=black!10] (0,1) circle (5pt);
\draw[fill=black!10] (1,0.4) circle (5pt);
\node[font=\fontsize{9}{9}] at (1,1.6) {$\textrm{F}$};
\node[font=\fontsize{9}{9}] at (0,1) {$\textrm{B}$};
\node[font=\fontsize{9}{9}] at (1,0.4) {$\textrm{F}$};
\end{tikzpicture} 
&
+
\vspace{1.8mm}
&
\begin{tikzpicture}[thick, scale=2]
\draw[decoration={markings, mark=at position 0.55 with {\arrow{stealth reversed}},mark=at position 0.08 with {\arrow{stealth}},mark=at position 0.96 with {\arrow{stealth}}},
        postaction={decorate}] (1,0) -- (1,2);
\draw[decoration={markings, mark=at position 0.55 with {\arrow{stealth}},mark=at position 0.08 with {\arrow{stealth reversed}},mark=at position 0.96 with {\arrow{stealth}}},
        postaction={decorate}] (1.36,0.186) -- (-0.36,1.21);
\draw[decoration={markings, mark=at position 0.55 with {\arrow{stealth reversed}},mark=at position 0.08 with {\arrow{stealth}},mark=at position 0.96 with {\arrow{stealth reversed}}},
        postaction={decorate}] (1.36,1.82) --(-0.36,0.78);
\draw[fill=black!10] (1,1.6) circle (5pt);
\draw[fill=black!10] (0,1) circle (5pt);
\draw[fill=black!10] (1,0.4) circle (5pt);
\node[font=\fontsize{9}{9}] at (1,1.6) {$\textrm{B}$};
\node[font=\fontsize{9}{9}] at (0,1) {$\textrm{S}$};
\node[font=\fontsize{9}{9}] at (1,0.4) {$\textrm{B}$};
\end{tikzpicture} 
\end{tabular}
\caption{\label{fig:Smatrix} The two factorization equations for the backward scattering $B$. On each line the scattering of an antiparticle (running downwards) with two particles (running upwards) decomposes into sequences of two-to-two scattering events controlled by the matrices $S, F, B$. Each line is associated to a rapidity that is carried either by a particle or an antiparticle according to the direction of the arrow.}
\end{figure}
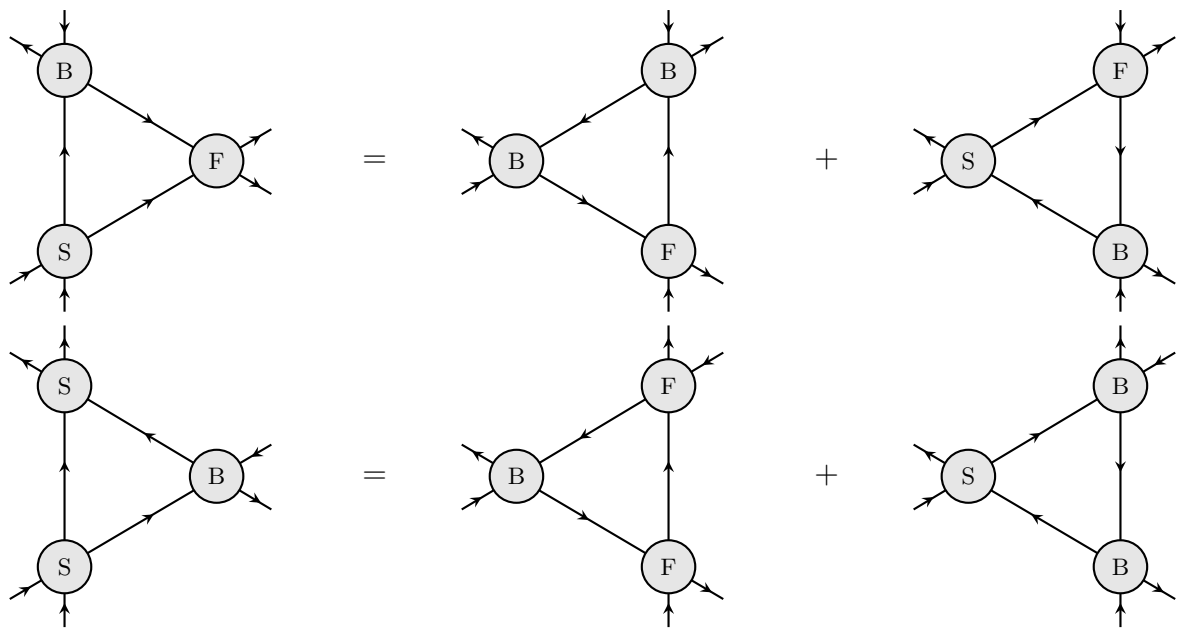

For theories with reflectionless scattering, when $B=0$, the relations \eqref{ut} are all we need. In all other cases one should also analyze the equations for the last two processes in \eqref{AssocFZ}. They relate the backward scattering $B$ to $S$ and $F$. We present these equations in a pictorial form in Figure \ref{fig:Smatrix}. 

\subsection{Minimal reflectionless S-matrix}

We will first turn to the minimal reflectionless S-matrix since it is of particular significance for the analysis carried out in this paper. By definition it is characterized by $r_1, r_2 = 0$. In this case the unitarity equations fix the value of the parameter $\nu$ in \eqref{ut} to
\beq
\nu = {2\pi \over \N}
\eeq
and lead to
\beq
t_{1}(\theta)t_{1}(-\theta) = 1\, , \qquad t_{1}(i\pi -\theta)t_{1}(i\pi+\theta) = {\theta^2 \over \theta^2+\nu^2}\, .
\eeq
The minimal S-matrix is a solution to these equations with the minimal set of singularities in the physical strip $\Im m\,  \theta \in [0, \pi]$. It belongs to the class II in the classification of \cite{Berg:1977dp}
\beq\label{MinSmatrix}
S(\theta) = -{\Gamma(1+x)\Gamma(\Delta -x) \over \Gamma(1-x)\Gamma(\Delta + x)}\, , \qquad t_1(\theta) = {\Gamma(\ft{1}{2}-x)\Gamma(\ft{1}{2}+\Delta +x) \over \Gamma(\ft{1}{2}+x)\Gamma(\ft{1}{2}+\Delta -x)}\,.
\eeq
In the above formulas $x = i\theta/2\pi$ and $\Delta = 1/\N$. All other amplitudes may be obtained by using \eqref{SphaseSym} and \eqref{ut}. This S-matrix has no singularity in the physical strip and thus the spinons do not form bound states. The overall sign, which is always left undetermined by the bootstrap program, was fixed by imposing that $S(\theta)=-1$. This condition is expected to hold true for bosonic models in general~\cite{Klassen:1992eq}. Here we simply observe that this is the correct choice for matching with the large $\N$ amplitudes of the $U(1)\times SU(\N)$ models considered in this paper.

The S-matrix~(\ref{MinSmatrix}) has two remarkable properties. The first one concerns the spinon scattering phase in the symmetric channel $S(\theta)$. It is easily seen to be identical up to an overall sign to the one found for the $SU(\N)$ chiral Gross-Neveu model~\cite{Koberle:1979ne} if  the substitution $\Delta \rightarrow 1-\Delta$ is performed. We made use of this relation when computing the free energy density in the single-spin ground state, see Section.~\ref{MMG}. The second remarkable feature is the scattering phase of a fundamental excitation with spinon-antispinon pair sharing the same rapidity.\footnote{This state can be viewed as an adjoint bound state at threshold.} This phase is simply given by the product $S(\theta)t_1(\theta)$ and is identical to the scattering factor in the symmetric channel of the $O(\N+2)$ sigma model \cite{Zamolodchikov:1978xm}. We did not make explicit use of this equivalence in this paper, but it would be relevant for the computation of the ground state energy density of the two equal spins solution mentioned at the end of the subsection~\ref{MMG}. 

Finally let us comment on the large rapidity expansion of  $S(\theta)$ and $t_1(\theta)$
\beq
S(\theta) \sim e^{-i\pi \Delta}\,, \qquad t_1(\theta) \sim e^{i\pi \Delta}\,, \qquad \theta \gg 1\, .
\eeq
This asymptotic behavior is indicating that the spinons behave as particles of  fractional Lorentz spin, quite similarly to the case of the chiral Gross-Neveu model~\cite{Koberle:1979ne}. To leading order at large $\N$ we have $\Delta \rightarrow 0$ and the spinons are boson as expected.

\subsection{S-matrix with backward scattering}

We turn back to scattering with reflection when $r_1(\theta), r_2(\theta) \neq 0$. To determine these extra factors we need to consider the Yang-Baxter equations for the two last processes in  (\ref{ut}). These are matrix equations that can be expanded over a basis of $SU(\N)$ covariant tensors. For generic values of $\N$ this leads to $5+5$ independent functional relations for the $S, F, B$ amplitudes. But not for $\N=2$. For this value of $\N$ only $4+4$ equations are available. This is due to the following identity between Kronecker symbols,
\beq
\delta_{ik}\delta_{jm}\delta_{ln} = \delta_{in}\delta_{jm}\delta_{kl} - \delta_{im}\delta_{jn}\delta_{kl} - \delta_{jk}\delta_{lm}\delta_{in} + \delta_{ik}\delta_{lm}\delta_{jn}+\delta_{jk}\delta_{ln}\delta_{im}\, ,
\eeq
which is valid \textit{only} when the indices run through two values. This may also be understood from the viewpoint of group theory. For the processes considered the $SU(\N)$ decomposition
\beq\label{NNbN}
N\otimes N \otimes \bar{N} = {(N+2)N(N-1) \over 2} \oplus {(N+1)N(N-2) \over 2}  \oplus N \oplus N\, ,
\eeq
applies only for $\N>2$. Each number on the r.h.s indicates the dimensions of the associated irreps and $N/\bar{N}$ in the l.h.s stand for the fundamental/antifundamental of $SU(\N)$. For $\N=2$ the irreps $\N$ and $\bar{\N}$ are isomorphic and the second channel in the r.h.s of (\ref{NNbN}) is absent. It is natural to expect less constraints in this case.

Let us now consider more closely these functional identities for the $S, F, B$ amplitudes. Two of them are valid for \textit{all} $\N$ 
\beq\label{SGeq}
\begin{aligned}
&r_1(\theta_{12})S(\theta_{13})r_1(\theta_{23}) + t_1(\theta_{12})r_1(\theta_{13})t_1(\theta_{23}) = S(\theta_{12})r_1(\theta_{13})S(\theta_{23})\, , \\
&t_1(\theta_{12})S(\theta_{13})r_1(\theta_{23}) + r_1(\theta_{12})r_1(\theta_{13})t_1(\theta_{23}) = S(\theta_{12})t_1(\theta_{13})r_1(\theta_{23})\, .
\end{aligned}
\eeq
After renaming
\beq
t_1(\theta) = S_{T}(\theta)\, , \qquad r_1(\theta) = S_{R}(\theta)\, ,
\eeq
they are revealed to be equivalent to the equations for the sine-Gordon theory~\cite{Zamolodchikov:1978xm}. This should not be surprising since the equations~(\ref{SGeq}) can be obtained by considering spinon-antispinon scattering in the $U(1)$ subsector, which is spanned by $Z_1$ and $\bar{Z}_{\N}$. The solution to the $U(1)$ factorized equations \eqref{SGeq} is known to be given by \cite{Zamolodchikov:1978xm}
\beq\label{StSr}
S(\theta) = -i{\sinh{(i\tilde{a}-a\theta)} \over \sin{\tilde{a}}}S_{R}(\theta)\, , \qquad S_{T}(\theta) = i{\sinh{a\theta} \over \sin{\tilde{a}}}S_{R}(\theta)\,.
\eeq
The parameters $a$ and $\tilde{a}$ are arbitrary, though the latter is defined modulo $2\pi$. This hints at the possibility of having  solutions with continuous parameters. To verify this one has to also solve the equations accounting for the $SU(\N)$ degrees of freedom. This is where the aforementioned difference between $\N>2$ and $\N=2$ plays a crucial role.

In the generic case $\N>2$, two of these additional equations are similar to (\ref{SGeq}) with the only difference being the substitution $S(\theta) = u_1(\theta)+u_1(\theta) \rightarrow u_2(\theta)-u_1(\theta)$. This however immediately excludes any continuous solutions since the ratio $u_1(\theta)/u_2(\theta)$ has been already constrained in (\ref{ut}) and is proportional to $\theta$. The consequence is that the sine-Gordon type expressions~(\ref{StSr}) become rational and the continuous parameters are lost. In other words the extra $SU(\N)$ equations spoil the freedom in (\ref{StSr}). Assuming $r_1, r_2, t_1, u_1 \neq 0$ and taking into account the remaining $SU(\N)$ equations, one finds only two possible solutions
\beq
t_1(\theta) = cu_1(\theta)\, , \qquad r_1(\theta) = u_2(\theta)\, , \qquad t_2(\theta) = r_2(\theta)\,,
\eeq
corresponding to $c=\pm1$. Unitarity and crossing yield then the additional relation
\beq
\nu = {\pi \over \N-c}\, .
\eeq 
The minimal solutions corresponding to $c=\pm 1$ are respectively given by the class III and IV solutions of \cite{Berg:1977dp}. The class III S-matrix is nothing else than the S-matrix for the $O(2\N)$ sigma model, while the class IV, to our knowledge, has not yet emerged as a scattering matrix of an integrable sigma model. Thus, as pointed out by Berg et al.~\cite{Berg:1977dp}, there is no room for a one-parameter family of solution if $\N> 2$. In particular, for generic values of $\N$ there is no interpolation between the reflectionless S-matrix and the $O(2\N)$ S-matrix, i.e. between type II and type III solutions in the classification of \cite{Berg:1977dp}.

Interestingly enough, this conclusion is invalid for $\N=2$. Of course all solutions found at generic $\N$ exist also at $\N=2$. But with less number of constraints we expect a larger class of solutions, perhaps even with continuous parameters. This is in fact what happens and we have checked explicitly that Wiegmann's S-matrix~\cite{Wiegmann:1985jt} solves all bootstrap equations for $\N=2$. The full solution reads
\beq\label{WiegmannSol}
S(\theta) = -S_{SU(2)}(\theta)S_{p}(\theta)\, ,
\eeq
with the sine-Gordon soliton scattering phase
\beq
S_p(\theta) = \exp{i\int_{0}^{\infty}{d\omega \over \omega} {\sin{(\omega\theta)}\sinh{(\pi\omega(\p-1)/2)} \over \cosh{(\pi\omega/2)}\sinh{(\pi\omega \p/2)}}}\, ,
\eeq
and the minimal $SU(2)$ scattering phase
\beq
S_{SU(2)}(\theta) = S_{p=\infty}(\theta) = {\Gamma(1+\ft{i\theta}{2\pi})\Gamma(\ft{1}{2}-\ft{i\theta}{2\pi}) \over \Gamma(1-\ft{i\theta}{2\pi})\Gamma(\ft{1}{2}+\ft{i\theta}{2\pi})}\, .
\eeq
All other amplitudes can be obtained by using \eqref{StSr} and \eqref{ut},  setting $a = 1/\p$, $\tilde{a} = a\pi$, $\nu = \pi$, and imposing crossing symmetry.  For $\p=\infty$ it is the $O(4) \simeq SU(2)\times SU(2)$ minimal solution while at $\p=1$ it becomes the minimal $U(2)$ reflectionless S-matrix interpolating between class II and III of \cite{Berg:1977dp}. The overall sign in~(\ref{WiegmannSol}) is chosen in accord with the generic $\p=1$ solution (\ref{MinSmatrix}).

We should mention here that there is a slight difference with the sine-Gordon S-matrix, in case of which the crossing symmetry requires that $\tilde{a} = a\pi + \pi$ in \eqref{StSr}. This is different from $\tilde{a} = a\pi$ for $\N=2$. The lack of the extra $\pi$ amounts to changing the sign of $S_{T}(\theta)$ which in turn interchanges sectors which are even and odd under the action of the charge conjugation $C$.  Mathematically, one has $S_{T} \pm S_{R} \rightarrow S_{T}\mp S_{R}$. This modification is harmless as far as the physical properties of the model are concerned, but is important for attributing correct $C$-parity to the bound states in the $\N=2$ theory.

\section{Counting argument : explicit calculation}\label{sec:CA}
In Section \ref{sec:first glimpse} we have provided a synopsis of the counting argument proposed by Goldschmidt and Witten \cite{Goldschmidt:1980wq}. We refer the reader to that paper for further details and explanations. Here we will apply this argument to the fermonic model. The following properties need to be observed when classifying the independent local operators.
\begin{itemize}
\item Taking derivatives of the kinematical constraint $\zb z=1$ and the Gauss law (\ref{BcFc}) allow to discard the operators $\bar{z}D^{n}_{\pm}z$ and their charge conjugates. 
\item String of derivatives of opposite helicity can always be broken up. This follows from the equations of motion~(\ref{eom}) and the commutator
\beq
[D_{+}, D_{-}] \sim (D_{-} \zb D_{+}z-D_{+}\zb D_{-}z)\, .
\eeq
\item Similarly the helicity of the fermion should have the same sign as the derivatives acting on it. For example, the operators $D_{\pm} \psi_\mp$ and their derivatives are forbidden.
\item The theory is invariant under charge conjugation. This property naturally extends to operators in both lists. 
\end{itemize}
Given the remarks above it is rather straightforward, if a little tedious, to compose lists A and B for the conservation law
\beq \label{Tmm}
\pa_{+} \left(T_{--}\right)^2 =0\,.
\eeq
We found that the set of anomalies consists of 16 elements
\beqa 
\nn
&&A=\Big\{ D^4_- \zb  D_+ z + c.c.\,,\,\,\, (D^2_- \zb D_- z)(D_- \zb D_+ z)+c.c.\,,\,\,\, (D^2_- \zb D_- z)(D_+ \zb D_- z)+c.c.\,,\,\,\,\\
\nn
&&\hspace{10mm} (D^2_- \zb D_+ z)(D_- \zb D_- z)+c.c.\,,\,\,\,
 \psi^*_- D^3_- \psi_- \psi^*_+ \psi_+ +c.c.\,,\,\,\, D_-\psi^*_- D^2_- \psi_- \psi^*_+ \psi_+ +c.c. \,,\,\,\, \\
\nn
&&\hspace{10mm} \partial_{-}(D_-\zb D_- z) \psi^*_- \psi_- \psi^*_+ \psi_+\,,\,\,\, D_-\zb D_- z\, \partial_{-}(\psi^*_- \psi_- )\psi^*_+ \psi_+\,,\,\,\, \\
\nn
&&\hspace{10mm} i(D_- \zb  D_+ z -c.c.)D_- \zb  D_- z\, \psi^*_- \psi_-\,,\,\,\, i (D^2_-\zb D_+ z +c.c.)(\psi^*_- D_- \psi_- -c.c.)\,,\,\,\,\\
\nn
&&\hspace{10mm} i(D^2_-\zb D_+ z - c.c.)\partial_{-}(\psi^*_- \psi_- ) \,,\,\,\, i(D^3_- \zb D_- z -c.c.) \psi^*_+ \psi_+ \,,\,\,\, \\
\nn
&&\hspace{10mm}  i(D_- \bar{z} D_+ z +c.c.) (\psi^*_- D^2_- \psi_- -c.c.)\,,\,\,\, i(D_- \bar{z} D_+ z -c.c.) \, (\psi^*_- D^2_- \psi_- +c.c.)\,,\,\,\, \\
\nn
&&\hspace{10mm} i(D_- \zb D_+ z-c.c.) D_- \psi^*_- D_- \psi_- \,,\,\,\, i(D^3_- \zb D_+ z -c.c.) \psi^*_- \psi_-\Big\}\, ,
\eeqa
where $c.c.$ denotes complex conjugation. The list of divergencies consists of two disjoint subsets
\beq
B=B_+ \cup B_-\,,
\eeq
with
\beqa
\nn
&&B_+=\partial_+\Big\{D ^3_- \zb  D_- z +c.c.\,,\,\,\,D ^2_- \zb  D^2_- z\,,\,\,\,i\psi^*_- D ^3_- \psi_- + c.c.\,,\,\,\,iD_- \psi^*_- D ^2_- \psi_- +c.c.\,, \\
\nn
&&\hspace{15mm}iD _- \zb D_- z (\psi^*_- D _- \psi_- - c.c.)\,,\,\,\,i(D ^2_- \zb D_- z-c.c.) \psi^*_-\psi_-\,,\,\,\,\\
\nn
&&\hspace{15mm}(D _- \zb D_- z)^2\,,\,\,\,\,D _- \psi^*_- D _- \psi_- \psi^*_- \psi_-  \Big\}\,,\\
\nn
&&B_-=\partial_-\Big\{D ^3_- \zb  D_+ z +c.c. \,,\,\,\,D _- \zb D_- z(D _- \zb D_+ z+c.c.) \,,\,\,\,(\psi^*_-D ^2_- \psi_-  +c.c.)\psi^*_+ \psi_+\\
\nn
&&\hspace{15mm}D _- \psi^*_- D _- \psi_- \psi^*_+ \psi_+\,,\,\,\,i(D ^2_- \bar{z} D_- z -c.c.) \psi^*_+ \psi_+\,,\,\,\, \\
\nn
&&\hspace{15mm}D _- \zb D_- z \,\psi^*_- \psi_- \psi^*_+\psi_+\,,\,\,\,i(D _+ \zb D_- z+c.c.)(\psi^*_-D _- \psi_- -c.c.)\,,\,\,\, \\
\nn
&&\hspace{15mm}i(D _+ \zb D_- z-c.c.)\partial_{-}(\psi^*_-\psi_-)\,,\,\,\,i(D ^2_- \bar{z} D_+ z -c.c.) \psi^*_- \psi_- \Big\}\,.
\eeqa
It contains 17 elements. We should be careful with the interpretation of the results though. Since there are more operators in the set B than in the set A, a particular combination of the former operators should vanish at quantum level. This identity may be easily derived by differentiating the conservation law for the stress-energy tensor
\beq
\pa_{+} T_{--} + \pa_{-} T_{+-}=0\,.
\eeq
Note that the above equation, contrary to the one obtained by setting $T_{+-}=0$, holds true also quantum-mechanically. Thus one element needs to be removed from the B-list. To check whether the conservation law \eqref{Tmm} may survive the quantization, we should also strip off the combination $\pa_+ (T_{--})^2$ or, equivalently, one of its constituent operators. This leaves us with 16 anomalies and only 15 divergencies, that is, with \textit{one} unmatched anomaly.  

The above counting is generic and is oblivious to the fact that at special values of $\N$ some operators may become redundant. This phenomenon might happen because our previous analysis does not guarantee that the kinematical constraint and Gauss law were exhausted. As it turns out, an exceptional relation among operators of the A-list may be found for $\N=2$, quite similarly to what we observed in Section~\ref{IBM}. This relation can be understood as a fermionic deformation of the geometrical identity valid for the  $CP^{1}$ model \cite{Goldschmidt:1980wq}. Classically it is expressed as
\beq\label{GW}
(D_{-}^{2}\bar{z}D_{-} z)(D_{-}\bar{z}D_{+}z) - (D_{-}^{2}\bar{z}D_{+} z)(D_{-}\bar{z}D_{-}z) + c.c. = \textrm{fermionic operators}\, ,
\eeq
where the r.h.s is a certain linear combination of (mixed) fermionic operators from the A-list, for example $\partial_{-}(D_-\zb D_- z) \psi^*_- \psi_- \psi^*_+ \psi_+$, etc. Irrespectively of its precise form, which is likely to be quantum corrected anyway, such a relation allows us to eliminate \textit{one} operator from the A-list. This is the sole modification and the B-list is valid as it stands. One thus infers that a non-trivial higher-spin conserved current should exist for $\N=2$. This mechanism was first observed for the $CP^{1}$ model \cite{Goldschmidt:1980wq}. Can we find relation similar to~(\ref{GW}) for $\N>2$? We believe this is not possible for the following reason. If such a relation existed it should be a fermionic deformation of an identity from the $CP^{\N-1}$ model. The latter however does not exist \cite{Goldschmidt:1980wq} and we expect the above counting to be correct for any $\N>2$.

Finally we stress that the number of unmatched anomalies is independent of the value of the Thirring coupling and the fermionic charge. This number also coincides with its counterpart for the $CP^{\N-1}$ sigma model~\cite{Goldschmidt:1980wq}. The important difference however is that for the fermionic extension we have \textit{one} continuous parameter at our disposal. Can one pick its value such that the coefficient of the unmatched anomaly is cancelled for $\N>2$? The answer seems to be positive as the results in other sections of this paper indicate.  
\bibliography{cpnpaper}
\bibliographystyle{nb}
\end{document}